\documentclass[lettersize,journal]{IEEEtran}
\usepackage{amsmath,amsfonts}
\usepackage{algorithmic}
\usepackage{array}
\usepackage[caption=false,font=normalsize,labelfont=sf,textfont=sf]{subfig}
\usepackage{textcomp}
\usepackage{stfloats}
\usepackage{url}
\usepackage{verbatim}
\usepackage{graphicx}
\def\BibTeX{{\rm B\kern-.05em{\sc i\kern-.025em b}\kern-.08em
    T\kern-.1667em\lower.7ex\hbox{E}\kern-.125emX}}
\usepackage{balance}
\usepackage{boondox-cal}
\usepackage{cite}
\usepackage{amssymb}
\usepackage{color}
\usepackage[ruled,linesnumbered]{algorithm2e}
\usepackage{diagbox}
\usepackage{makecell}

\begin{document}
	
\title{Log Density Spectrum Based Decoding of Polar Codes for Channels with Strong Phase Noise}
\author{Yinuo Mei, and Daiming Qu
	\thanks{This work was supported in part by the National Science Foundation of China (No. 62331010). (Corresponding author: Daiming
		Qu.)
		
		The authors are with the School of Electronic Information and
		Communications, Huazhong University of Science and Technology,
		Wuhan 430074, China. (email: D202280923@hust.edu.cn; qudaiming@hust.edu.cn).}}

\maketitle

\begin{abstract}
A decoding method is proposed for polar codes
under channels with strong phase noise. In contrast to traditional SC
decoding, where recursions are performed on scalar soft information such
as probabilities, the proposed decoding recurses over probability
density functions defined on the phase-noise variable. To make the
density-level recursion computationally tractable, we represent each
log-density by its spectrum, and derive approximate formulas for the
spectrum recursion, therefore, the proposed method is termed log-density
spectrum recursion (LDSR). Numerical results are given under strong
phase noise and AWGN, showing that a prominent superiority of error
performance is achieved by LDSR in comparison with traditional SC and
Kalman-based methods.
\end{abstract}


\section{Introduction}

\IEEEPARstart{I}{n} modern high-speed communication systems, phase noise represents a
critical bottleneck for achieving high spectral efficiency and reliable
coherent detection. Primarily originating from the inherent
instabilities of local oscillators and lasers, phase noise manifests as
random fluctuations in the carrier phase, leading to constellation
rotation and increased decision errors at the receiver \cite{ref:PN}. In
high-frequency and long-haul transmission regimes, including sub-THz
bands, satellite links, optical transport networks, etc., phase noise
can be particularly strong and may severely degrade transmission
performance \cite{ref:sub-THz,ref:satellite,ref:optical}. Beyond hardware-induced fluctuations,
phase impairments are severely exacerbated in high-mobility scenarios,
such as high-speed railways or UAV networks. In these environments,
rapid relative motion introduces significant Doppler shifts; any
residual frequency offset after initial compensation linearly
accumulates over time, resulting in a burgeoning phase drift
\cite{ref:Doppler1,ref:Doppler2}. This cumulative phase error, coupled with stochastic
oscillator noise, creates a highly non-stationary channel environment
that can quickly outpace traditional tracking loops. These impairments
significantly degrade signal-to-noise ratios (SNR), increase bit-error
rates (BER), and limit the deployment of high-order modulation schemes.
Consequently, developing robust phase-error mitigation techniques that
can jointly track random jitter and compensate for mobility-induced
Doppler accumulation is essential for the next generation of resilient
communication links.

To tackle phase noise, various approaches have been proposed, among
which Kalman filter/smoother-based methods are classical and widely
adopted. However, the application of Kalman filters/smoothers necessitates a priori knowledge of the transmitted symbols, which is not always available at the receiver. Consequently, pilot symbols are commonly
inserted into the interior of the data frame, namely scattered across the data frame, to facilitate phase noise estimation \cite{ref:Kalman Pilot1,ref:Kalman Pilot2,ref:Kalman Pilot3,ref:Kalman Pilot4,ref:Kalman Pilot5}, inevitably
incurring pilot overhead and reducing spectral efficiency. To avoid this
overhead, the pilot-free iterative detection and decoding scheme has been
developed as an effective alternative, where a Kalman filter/smoother
is embedded within the turbo-like iteration loop between the detector
and the decoder \cite{ref:Kalman Soft}. Nevertheless, within the iterative process, the Kalman filter/smoother treats phase noise estimates as hard decisions, which inevitably discards soft
information, thereby degrading error-correction performance. Especially under strong phase noise, the loss of information would cause a severe performance degradation, which poses a critical barrier to the Kalman-based iterative scheme. Therefore, it is expected to fully exploit the soft information of phase noise, rather than make decisions on it. To enable the exploitation of soft information of phase noise, a joint detection and decoding framework serves as a highly competitive option, which directly incorporates a
dedicated probabilistic model of phase noise into the decoder.

While joint detection and decoding is essential to combat strong phase
noise, its practical effectiveness heavily relies on the underlying
channel coding scheme. Among various channel codes, polar codes are characterized by an explicit recursive coding structure
\cite{ref:Polar}, accordingly, their successive
cancellation (SC)-based decoding scheduling \cite{ref:Polar,ref:SCL} provides a natural
framework for integrating soft information of phase noise with code
structure, which is to be further explained in Section III. In addition, owing
to their capacity-achieving property \cite{ref:Cap,ref:Polar}, polar codes have emerged as an attractive candidate for modern
communications \cite{ref:polarApp1,ref:polarApp2,ref:polarApp3}. Motivated by
these appealing features, in this paper, we investigate polar codes, and
address the decoding of polar codes in the presence of strong phase
noise. Besides, BPSK-modulated Gaussian channel is a canonical setup for
polar codes \cite{ref:SCL}. Owing to its simplicity and tractability, this
widely used benchmark not only enables fair comparisons with established
polar coding results, but also cleanly highlights the degradation
induced purely by phase fluctuations. Therefore, throughout this paper,
we focus on BPSK modulation over a channel subject to both phase noise
and additive white Gaussian noise (AWGN).

The rest of this paper is organized as follows. Section II gives preliminaries of phase noise and polar codes. In Section III, the polar decoding under phase noise channel is discussed, which does not yet have a closed-form representation. Then its closed-form approximation is developed in Section IV, leading to a tractable decoding algorithm. Section V analyzes the complexity of the proposed decoding algorithm. The simulation results are presented in Section VI. Finally, Section VII serves as a conclusion of this research.

Notations: The sets of real numbers, positive real numbers, integers, natural numbers, positive integers, and \(\{0,1\}\) are denoted by \(\mathbb{R}\), \(\mathbb{R}^+\), \(\mathbb{Z}\), \(\mathbb{N}\), \(\mathbb{N}^*\), \(\mathbb{B}\), respectively. \(c_{[k,m]}\) denotes the \((k,m)\)-th element of infinite matrix \(\boldsymbol{c}\). \(z^*\) denotes the conjugate of \(z\). \(\boldsymbol{c}^*\) denotes the conjugate of \(\boldsymbol{c}\), i.e., \((\boldsymbol{c}^*)_{[k,m]}:=\left(c_{[k,m]}\right)^*\), and \(\boldsymbol{c}^{*-}\) denotes the conjugate reversal of \(\boldsymbol{c}\), i.e., \((\boldsymbol{c}^{*-})_{[k,m]}:=\left(c_{[-k,-m]}\right)^*\). \(\boldsymbol{x}*\boldsymbol{y}\) refers to the convolution of infinite matrices \(\boldsymbol{x}\) and \(\boldsymbol{y}\), i.e., \((\boldsymbol{x}*\boldsymbol{y})_{[k,m]}:=\sum_{l=-\infty}^\infty \sum_{n=-\infty}^\infty x_{[l,n]}y_{[k-l,m-n]}\). \(\mathrm{supp}(\boldsymbol{c})\) refers to the support set of \(\boldsymbol{c}\), namely the set of indices at which \(\boldsymbol{c}\) takes non-zero values, i.e., \(\mathrm{supp}(\boldsymbol{c}):=\left\{(k,m)\middle|c_{[k,m]}\neq0\right\}\). \(\boldsymbol{\delta}_{p,q}\) is an infinite matrix defined by \(\delta_{p,q,\lbrack k,m\rbrack} := \left\{ \begin{aligned}
	& 1,k = p, m = q \\
	& 0,\mathrm{otherwise}
\end{aligned} \right. \). Moreover, \(\boldsymbol{\delta}_{p,-}\) is an infinite matrix defined by
\(\delta_{p,-,\lbrack k,m\rbrack} := \left\{ \begin{aligned}
	& 1,k = p\\
	& 0,\mathrm{otherwise}
\end{aligned} \right. \), and \(\boldsymbol{\delta}_{-,q}\) is an infinite matrix defined by
\(\delta_{-,q,\lbrack k,m\rbrack} := \left\{ \begin{aligned}
& 1,m = q\\
& 0,\mathrm{otherwise}
\end{aligned} \right. \).

\section{Preliminaries}

This paper focuses on polar decoding under channels with phase noise, therefore the
preliminaries of phase noise and polar codes are presented below.

\subsection{Phase Noise}

A commonly used model for phase noise is the discrete-time Wiener
process \cite{ref:PN,ref:sub-THz,ref:satellite,ref:optical}, in which the phase noise evolves
according to:
\begin{align} \label{eq:PN}
	\Phi_{n+1} = \Phi_{n} + E_{n+1}
\end{align}
where \(E_{n}\sim \mathcal{N}\left( 0,\sigma_{\phi}^{2} \right)\),
\(\sigma_{\phi}^{2}\) is termed the phase transition variance, and \(n\)
represents the symbol index,
\(n \in \left\{ 0,1,\ldots,N - 1 \right\}\), \(N\) is the code length.
\(E_{0}^{N - 1}\) is i.i.d. Evidently, the stochastic process of random
phase noise \(\Phi_{0}^{N-1}\) is a Markov process:
\begin{align}
	p_{\Phi_{n+1}|\Phi_{0}^{n}}\left( \phi_{n+1} \middle| \phi_{0}^{n} \right) = p_{\Phi_{n+1}|\Phi_{n}}(\phi_{n+1}|\phi_{n})
\end{align}
the latter of which is expressed by
{\small
\begin{align} \label{eq:PNden}
	p_{\Phi_{n+1}|\Phi_{n}}\left( \phi_{n+1} \middle| \phi_{n} \right) = \frac{1}{\sqrt{2\pi}\sigma_{\phi}}\exp\left\{ - \frac{\left( \phi_{n+1} - \phi_{n} \right)^{2}}{2\sigma_{\phi}^{2}} \right\}
\end{align}
}

\subsection{Polar Codes}

Polar codes are a class of linear block codes. Encoding of polar codes
with code length \(N\) and information bit number \(K\), namely code
rate \(R = \frac{K}{N}\), includes assigning information bits
\(c_{0}^{K - 1}\) to an \emph{N}-length bit sequence \(u_{0}^{N - 1}\),
and mapping source bits \(u_{0}^{N - 1}\) to final codeword
\(x_{0}^{N - 1}\) with a linear operator, which can be represented as a
matrix \(G_{N}\) termed generator matrix. In expressions, codeword
\(x_{0}^{N - 1} = u_{0}^{N - 1}G_{N}\), where \(u_{0}^{N - 1}\) is a
combination of information bits \(c_{0}^{K - 1}\) and frozen bits zeros,
namely
\(u_{\mathcal{A}} = c_{0}^{K - 1},u_{\mathcal{A}^{\complement}} = \lbrack 0,\ldots,0\rbrack\).
Here \(\mathcal{A}\) is the information index set, namely the index set allocated to
information bits, contrarily bits at indices
\(\mathcal{A}^{\complement} = \left\{ 0,1,\ldots,N - 1 \right\} \backslash \mathcal{A}\) in \(u_{0}^{N - 1}\) are zeros. Namely,
\(x_0^{N-1}=u_{\mathcal{A}} G_N(\mathcal{A})\), where \(G_N\left( \mathcal{A} \right)\)
refers to the submatrix of \(G_N\) composed of its rows indexed in
\(\mathcal{A}\). Generator matrix of polar codes with code
length \(N = 2^{n},n \in \mathbb{N}^{*}\) is
\(G_{N} = B_{N}F^{\otimes n}=F^{\otimes n}B_{N}\), where \(B_{N}\) is a bit-reversal
permutation matrix, and \(F = \begin{bmatrix}
	1 & 0 \\
	1 & 1
\end{bmatrix}\) is termed polar kernel, besides, operator \(\otimes n\)
means \emph{n}-th Kronecker power \cite{ref:Polar}.

Then the codeword \(x_0^{N-1}\) is transmitted through a memoryless channel \(W^N\), which represents the \(N\) uses of binary-input channel \(W\), and the output of \(W^N\) is \(y_0^{N-1}\). While decoding, polar code utilizes received signal and former source
bits to estimate each source bit, specifically, it constructs \(N\)
channels for every source bit, termed sub-channels, the \emph{i}-th of
which is
\(U_{i} \mapsto Y_{0}^{N - 1}U_{0}^{i - 1},i \in \left\{ 0,1,\ldots,N - 1 \right\}\).
So the transition probability of the \emph{i}-th sub-channel is
\(W_{N}^{(i)}\left( y_{0}^{N - 1},u_{0}^{i - 1} \middle| u_{i} \right)\).
Due to the recursive property of generator matrix \(G_{N}\), on
memoryless channels, the recursive expression
between transition probability on adjacent layers is derived as \cite{ref:Polar}

{\footnotesize
	\begin{align} \label{eq:rece_trd}
		&W_{2N}^{(2i)}\left( y_{0}^{2N - 1},u_{0}^{2i - 1} \middle| u_{2i} \right)= \nonumber \\
		&\frac{1}{2}\sum_{u_{2i + 1}\in\mathbb{B}}^{}{W_{N}^{(i)}\left( y_{0}^{N - 1},v_{0,0}^{i - 1} \middle| u_{2i} + u_{2i + 1} \right)W_{N}^{(i)}\left( y_{N}^{2N - 1},v_{1,0}^{i - 1} \middle| u_{2i + 1} \right)}
	\end{align}
	
	\begin{align} \label{eq:reco_trd}
		&W_{2N}^{(2i + 1)}\left( y_{0}^{2N - 1},u_{0}^{2i} \middle| u_{2i + 1} \right)= \nonumber \\
		&\frac{1}{2}W_{N}^{(i)}\left( y_{0}^{N - 1},v_{0,0}^{i - 1} \middle| u_{2i} + u_{2i + 1} \right)W_{N}^{(i)}\left( y_{N}^{2N - 1},v_{1,0}^{i - 1} \middle| u_{2i + 1} \right)
	\end{align}
}where \(v_{0,i} \mathord{=} u_{2i} + u_{2i + 1},v_{1,i} \mathord{=} u_{2i + 1},\forall i \in \left\{ 0,1,\ldots,N - 1 \right\}\).

The fundamental decoding algorithm of polar code is successive
cancellation (SC) \cite{ref:Polar}. SC estimates each source bit \(u_{i}\) by
received signal \(y_{0}^{N - 1}\) and former estimated source
bits \({\widehat{u}}_{0}^{i - 1}\), and estimation at index \(i\) viz.
\({\widehat{u}}_{i}\) is exploited along with former estimations
\({\widehat{u}}_{0}^{i - 1}\) in subsequent decoding, as following
expression indicates:
\begin{align}
	{\widehat{u}}_{i} = 
	\begin{cases}
		\arg{\max_{u_{i} \in \mathbb{B}}{W_{N}^{(i)}\left( y_{0}^{N - 1},{\widehat{u}}_{0}^{i - 1} \middle| u_{i} \right)}} &,i \in \mathcal{A} \\
		0 &,i \in \mathcal{A}^{\complement}
	\end{cases}
\end{align}

Moreover, I. Tal and A. Vardy proposed a list decoding algorithm, which
saves multiple decoding paths during decoding, each of which records an
estimated bit sequence. Such algorithm is successive cancellation list (SCL) \cite{ref:SCL}. Considerable
performance and acceptable time and space complexity render it a practical
decoding algorithm for polar codes.

\section{Polar Decoding under Phase Noise Channel}

Polar codes are originally designed for memoryless channels.
Accordingly, traditional polar decoding method under memoryless
channels, i.e., SC is proposed, which relies on scalar probability
recursions governed by (\ref{eq:rece_trd}) and (\ref{eq:reco_trd}) \cite{ref:Polar}. However, as detailed in
Section II-A, phase noise exhibits strong temporal dependence across
adjacent samples, thereby introducing memory into the channel. This
channel memory invalidates the underlying assumption of standard SC
decoding, motivating the development of a dedicated polar decoding
algorithm tailored for phase noise channels.

\subsection{Proposed Decoding by Recursions over Log-Densities}

\begin{figure*}[!t]
	\centering
	\includegraphics[width=5in]{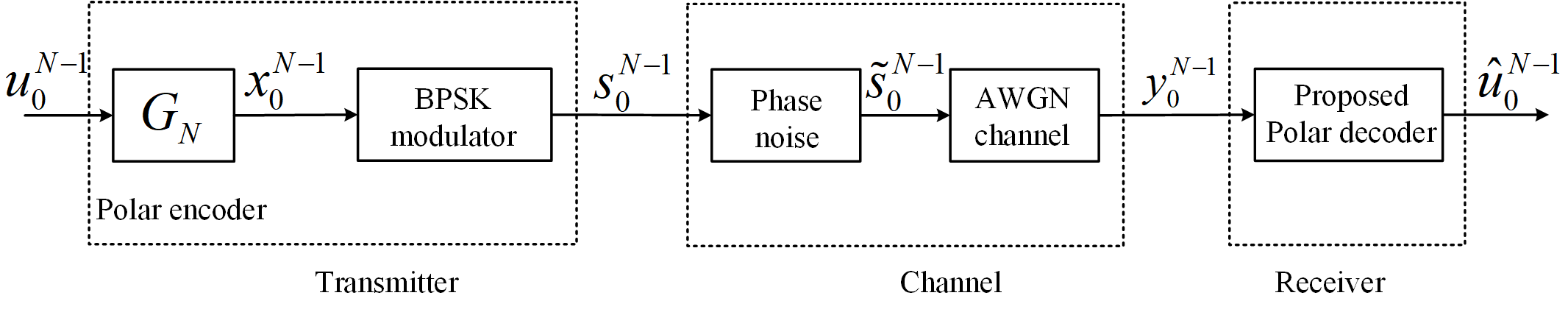}
	\caption{The system model.}
	\label{fig:sys}
\end{figure*}

The system model is illustrated in Fig.~\ref{fig:sys}. Under BPSK modulation, the
symbol sequence is
\(S_{0}^{N - 1} = e^{j\pi X_{0}^{N - 1}} = e^{j\pi U_{0}^{N - 1}G_{N}}\).
Through phase noise \(\Phi_{0}^{N - 1}\), the phase-shifted symbol
sequence
\({\widetilde{S}}_{0}^{N - 1} := S_{0}^{N - 1}e^{j\Phi_{0}^{N - 1}}\)
satisfies that
\begin{align}
	{\widetilde{S}}_{0}^{N - 1} = e^{j\pi U_{0}^{N - 1}G_{N}}e^{j\Phi_{0}^{N - 1}} = e^{j\left( \pi U_{0}^{N - 1}G_{N} + \Phi_{0}^{N - 1} \right)}
\end{align}
where \(\Phi_{0}^{N-1}\) is a Wiener process described in Section II-A.
Therefore, the channel is
\begin{align} \label{eq:Sys}
	Y_{0}^{N - 1} = e^{j\left( \pi U_{0}^{N - 1}G_{N} + \Phi_{0}^{N - 1} \right)} + \Delta_{0}^{N - 1}
\end{align}
where \(\Delta_{0}^{N - 1}\) is the AWGN. Within the overall channel
\(S_{0}^{N - 1} \mapsto Y_{0}^{N - 1}\), the phase-shifting
\(S_{0}^{N - 1} \mapsto {\widetilde{S}}_{0}^{N - 1}\) can be modeled as
a component-channel, termed phase-shifting component-channel. Namely,
the overall channel \(S_{0}^{N - 1} \mapsto Y_{0}^{N - 1}\) can be
decomposed to a phase-shifting component-channel
\(S_{0}^{N - 1} \mapsto {\widetilde{S}}_{0}^{N - 1}\) and an AWGN
channel \({\widetilde{S}}_{0}^{N - 1} \mapsto Y_{0}^{N - 1}\).

According to this system model, we have equations (\ref{eq:rece})(\ref{eq:reco}).
{\footnotesize
{\color{black}
\begin{align} \label{eq:rece}
	&p_{Y_{0}^{2N - 1}U_{0}^{2i - 1}\Phi_{2N}|U_{2i}\Phi_{0}}\left( y_{0}^{2N - 1},u_{0}^{2i - 1},\phi_{2N} \middle| u_{2i},\phi_{0} \right)= \nonumber \\
	& \frac{1}{2}\sum_{u_{2i + 1} \in \mathbb{B}}^{}\int_{0}^{2\pi}p_{Y_{0}^{N - 1}V_{0,0}^{i - 1}\Phi_{N}|V_{0,i}\Phi_{0}}\left( y_{0}^{N - 1},v_{0,0}^{i - 1},\phi_{N} \middle| u_{2i} + u_{2i + 1},\phi_{0} \right) \nonumber \\
	&\cdot p_{Y_{N}^{2N - 1}V_{1,0}^{i - 1}\Phi_{2N}|V_{1,i}\Phi_{N}}\left( y_{N}^{2N - 1},v_{1,0}^{i - 1},\phi_{2N} \middle| u_{2i + 1},\phi_{N} \right)\mathrm{d}\phi_{N}
\end{align}
\begin{align} \label{eq:reco}
	&p_{Y_{0}^{2N - 1}U_{0}^{2i}\Phi_{2N}|U_{2i + 1}\Phi_{0}}\left( y_{0}^{2N - 1},u_{0}^{2i},\phi_{2N} \middle| u_{2i + 1},\phi_{0} \right) \nonumber \\
	&= \frac{1}{2}\int_{0}^{2\pi} p_{Y_{0}^{N - 1}V_{0,0}^{i - 1}\Phi_{N}|V_{0,i}\Phi_{0}}\left( y_{0}^{N - 1},v_{0,0}^{i - 1},\phi_{N} \middle| u_{2i} + u_{2i + 1},\phi_{0} \right) \nonumber \\ 
	&\cdot p_{Y_{N}^{2N - 1}V_{1,0}^{i - 1}\Phi_{2N}|V_{1,i}\Phi_{N}}\left( y_{N}^{2N - 1},v_{1,0}^{i - 1},\phi_{2N} \middle| u_{2i + 1},\phi_{N} \right)\mathrm{d}\phi_{N}
\end{align}
}}where \(V_{0,i} = U_{2i} + U_{2i + 1},V_{1,i} = U_{2i + 1},\forall i \in \left\{ 0,\ldots,N - 1 \right\}\).

The proof of (\ref{eq:rece})(\ref{eq:reco}) is in Appendix-A.

Define log-density
{\small
\begin{align}
	&L_{N}^{(i)}\left( y_{0}^{N - 1},u_{0}^{i - 1},\phi_{N} \middle| u_{i},\phi_{0} \right) \nonumber \\
	&:= \ln{p_{Y_{0}^{N - 1}U_{0}^{i - 1}\Phi_{N}|U_{i}\Phi_{0}}\left( y_{0}^{N - 1},u_{0}^{i - 1},\phi_{N} \middle| u_{i},\phi_{0} \right)}
\end{align}
}then (\ref{eq:rece})(\ref{eq:reco}) could be reformulated as
{\small
\begin{align} \label{eq:receLDv}
	&L_{2N}^{(2i)}\left( y_{0}^{2N - 1},u_{0}^{2i - 1},\phi_{2N} \middle| u_{2i},\phi_{0} \right) \nonumber \\
	&= \frac{1}{2}\sum_{u_{2i + 1}\in\mathbb{ B}}^{} \int_{0}^{2\pi} \exp \left\{ L_{N}^{(i)}\left( y_{0}^{N - 1},v_{0,0}^{i - 1},\phi_{N} \middle| u_{2i} + u_{2i + 1},\phi_{0} \right) \right. \nonumber \\
	& \left. + L_{N}^{(i)}\left( y_{N}^{2N - 1},v_{1,0}^{i - 1},\phi_{2N} \middle| u_{2i + 1},\phi_{N} \right) \right\} \mathrm{d} \phi_{N}
\end{align}
\begin{align} \label{eq:recoLDv}
	&L_{2N}^{(2i + 1)}\left( y_{0}^{2N - 1},u_{0}^{2i},\phi_{2N} \middle| u_{2i + 1},\phi_{0} \right) \nonumber \\
	&= \frac{1}{2}\int_{0}^{2\pi}\exp \left\{ L_{N}^{(i)}\left( y_{0}^{N - 1},v_{0,0}^{i - 1},\phi_{N} \middle| u_{2i} + u_{2i + 1},\phi_{0} \right) \right. \nonumber \\
	& \left. + L_{N}^{(i)}\left( y_{N}^{2N - 1},v_{1,0}^{i - 1},\phi_{2N} \middle| u_{2i + 1},\phi_{N} \right) \right\} \mathrm{d} \phi_{N}
\end{align}
}

For brevity, under fixed \(y_{0}^{N - 1}\) and \(u_{0}^{i}\), we denote
the log-density function w.r.t. the phase noise \(\phi_{0},\phi_{N}\) by
\(\mathcal{L}_{N}^{(i)}\left( y_{0}^{N - 1},u_{0}^{i - 1} \middle| u_{i} \right)\),
i.e.,
\(\mathcal{L}_{N}^{(i)}\left( y_{0}^{N - 1},u_{0}^{i - 1} \middle| u_{i} \right) \in C\left( \lbrack 0,2\pi)^{2} \right)\),
where \(C\left( \lbrack 0,2\pi)^{2} \right)\) refers to the set of
continuous functions defined on \(\lbrack 0,2\pi)^{2}\), and
{\small
\begin{align}
	\mathcal{L}_{N}^{(i)}\left( y_{0}^{N - 1},u_{0}^{i - 1} \middle| u_{i} \right)|_{\phi_{0},\phi_{N}} := L_{N}^{(i)}\left( y_{0}^{N - 1},u_{0}^{i - 1},\phi_{N} \middle| u_{i},\phi_{0} \right)
\end{align}
}

To keep the notations compact, for
\(f_{0},f_{1} \in C\left( \lbrack 0,2\pi)^{2} \right)\), we define operators \(\odot\) and \(\oplus\) by
{\small
\begin{align} \label{eq:def_odot}
	\left( f_{0} \odot f_{1} \right)|_{\phi_{0},\phi_{2}} := \ln\left( \frac{1}{2}\int_{0}^{2\pi}{e^{f_{0}\left( \phi_{0},\phi_{1} \right) + f_{1}\left( \phi_{1},\phi_{2} \right)}\mathrm{d}\phi_{1}} \right)
\end{align}
\begin{align} \label{eq:def_oplus}
	\left( f_{0} \oplus f_{1} \right)|_{\phi_{0},\phi_{1}} := \ln\left( e^{f_{0}\left( \phi_{0},\phi_{1} \right)} + e^{f_{1}\left( \phi_{0},\phi_{1} \right)} \right)
\end{align}
}
then (\ref{eq:receLDv})(\ref{eq:recoLDv}) could be reformulated as
{\small
\begin{align} \label{eq:receLD}
	&\mathcal{L}_{2N}^{(2i)}\left( y_{0}^{2N - 1},u_{0}^{2i - 1} \middle| u_{2i} \right) \nonumber \\
	&= \underset{u_{2i + 1}\in\mathbb{ B}}{\oplus} \left\lbrack \mathcal{L}_{N}^{(i)}\left( y_{0}^{N - 1},v_{0,0}^{i - 1} \middle| u_{2i} + u_{2i + 1} \right) \right. \nonumber \\
	&\odot \left. \mathcal{L}_{N}^{(i)}\left( y_{N}^{2N - 1},v_{1,0}^{i - 1} \middle| u_{2i + 1} \right) \right\rbrack
\end{align}
\begin{align} \label{eq:recoLD}
	&\mathcal{L}_{2N}^{(2i + 1)}\left( y_{0}^{2N - 1},u_{0}^{2i} \middle| u_{2i + 1} \right) \nonumber \\
	&= \mathcal{L}_{N}^{(i)}\left( y_{0}^{N - 1},v_{0,0}^{i - 1} \middle| u_{2i} + u_{2i + 1} \right) \odot \mathcal{L}_{N}^{(i)}\left( y_{N}^{2N - 1},v_{1,0}^{i - 1} \middle| u_{2i + 1} \right)
\end{align}
}

In principle, by applying (\ref{eq:receLD})(\ref{eq:recoLD}) recursively, and finally applying
\(p_{Y_{0}^{N - 1}U_{0}^{N - 1}|U_{i}}\left( y_{0}^{N - 1},u_{0}^{i - 1} \middle| u_{i} \right) = \int_{0}^{2\pi}{\exp\left\{ L_{N}^{(i)}\left( y_{0}^{N - 1},u_{0}^{i - 1},\phi_{N} \middle| u_{i},0 \right) \right\} \mathrm{d}\phi_{N}}\),
we could obtain the transition probability of each subchannel
\(U_{i} \mapsto Y_{0}^{N - 1}U_{0}^{i - 1}\), and determine each source
bit \(u_{i}\) according to the transition probability. In this sense,
(\ref{eq:receLD})(\ref{eq:recoLD}) could be regarded as polar decoding under phase noise channel.

However, a continuous density itself cannot be stored and calculated by
an algorithm or program, thereby not tractable. Therefore, it is
expected to represent the continuous log-density
\(\mathcal{L}_{N}^{(i)}\left( y_{0}^{N - 1},u_{0}^{i - 1} \middle| u_{i} \right)\)
by a discrete matrix, i.e., to discretize log-densities, and develop a tractable algorithm of its
recursion.

\subsection{Discretization of Log-Densities by Spectral Representation}

According to two-dimensional continuous-time discrete Fourier
transformation (2D-CTDFT), for an arbitrary continuous function \(f\)
defined on \(\lbrack 0,2\pi)^{2}\), its spectrum
\(\boldsymbol{c}\)
is defined by
\begin{align}
	c_{[k,m]} = \int_{0}^{2\pi}{\int_{0}^{2\pi}{f\left( \phi_{0},\phi_{1} \right)e^{- jk\phi_{0}}e^{- jm\phi_{1}}}\mathrm{d}\phi_{0}}\mathrm{d}\phi_{1}
\end{align}
and conversely, there is
\begin{align}
	f\left( \phi_{0},\phi_{1} \right) = \sum_{k,m\in\mathbb{ Z}}^{}{c_{[k,m]}e^{jk\phi_{0}}e^{jm\phi_{1}}}
\end{align}
namely 2D-CTDFT is invertible, i.e., a function is
one-to-one mapped to a spectrum. Therefore, a continuous function
defined on \(\lbrack 0,2\pi)^{2}\) could be represented by its spectrum,
the latter being discrete. Although the spectrum is an infinite
matrix, assuming that the support set of it is finite, it could be
represented by a finite matrix.

Denote the 2D-CTDFT by \(\mathcal{F}\), i.e.,
\(\boldsymbol{c}=\mathcal{F}\lbrack f\rbrack\), \(f = \mathcal{F}^{- 1}\left\lbrack \boldsymbol{c} \right\rbrack\),
and define the log-density spectrum by
{\small
\begin{align}
	\boldsymbol{c}_{N}^{(i)}\left( y_{0}^{N - 1},u_{0}^{i - 1} \middle| u_{i} \right) := \mathcal{F}\left\lbrack \mathcal{L}_{N}^{(i)}\left( y_{0}^{N - 1},u_{0}^{i - 1} \middle| u_{i} \right) \right\rbrack
\end{align}
}then (\ref{eq:receLD})(\ref{eq:recoLD}) could be represented by the spectrum as follows.
{\footnotesize
\begin{align}
	&\boldsymbol{c}_{2N}^{(2i)}\left( y_{0}^{2N - 1},u_{0}^{2i - 1} \middle| u_{2i} \right) \nonumber \\
	&= \mathcal{F}\left\lbrack \underset{u_{2i + 1}\in\mathbb{ B}}{\oplus}\left\lbrack \mathcal{F}^{- 1}\left\lbrack \boldsymbol{c}_{N}^{(i)}\left( y_{0}^{N - 1},v_{0,0}^{i - 1} \middle| u_{2i} + u_{2i + 1} \right) \right\rbrack \right. \right. \nonumber \\
	&\left. \left. \odot \mathcal{F}^{- 1}\left\lbrack \boldsymbol{c}_{N}^{(i)}\left( y_{N}^{2N - 1},v_{1,0}^{i - 1} \middle| u_{2i + 1} \right) \right\rbrack \right\rbrack \right\rbrack
\end{align}
\begin{align}
	&\boldsymbol{c}_{2N}^{(2i + 1)}\left( y_{0}^{2N - 1},u_{0}^{2i} \middle| u_{2i + 1} \right) \nonumber \\
	&=\mathcal{F}\left\lbrack \mathcal{F}^{- 1}\left\lbrack \boldsymbol{c}_{N}^{(i)}\left( y_{0}^{N - 1},v_{0,0}^{i - 1} \middle| u_{2i} + u_{2i + 1} \right) \right\rbrack \right. \nonumber \\
	&\left. \odot \mathcal{F}^{- 1}\left\lbrack \boldsymbol{c}_{N}^{(i)}\left( y_{N}^{2N - 1},v_{1,0}^{i - 1} \middle| u_{2i + 1} \right) \right\rbrack \right\rbrack
\end{align}
}

Define operators \(\boxdot\) and \(\boxplus\) by
{\small
\begin{align} \label{eq:def_boxdot}
	\boldsymbol{c}_{1} \boxdot \boldsymbol{c}_{2}:= \mathcal{F}\left\lbrack \mathcal{F}^{- 1}\left\lbrack \boldsymbol{c}_{1} \right\rbrack \odot \mathcal{F}^{- 1}\left\lbrack \boldsymbol{c}_{2} \right\rbrack \right\rbrack
\end{align}
\begin{align} \label{eq:def_boxplus}
	\boldsymbol{c}_{1} \boxplus \boldsymbol{c}_{2}:= \mathcal{F}\left\lbrack \mathcal{F}^{- 1}\left\lbrack \boldsymbol{c}_{1} \right\rbrack \oplus \mathcal{F}^{- 1}\left\lbrack \boldsymbol{c}_{2} \right\rbrack \right\rbrack
\end{align}
}then above formulas are abbreviated to
{\small
{\color{black}
\begin{align} \label{eq:receSp}
	&\boldsymbol{c}_{2N}^{(2i)}\left( y_{0}^{2N - 1},u_{0}^{2i - 1} \middle| u_{2i} \right) \nonumber \\
	&= \underset{u_{2i + 1}\in\mathbb{ B}}{\boxplus}\left\lbrack \boldsymbol{c}_{N}^{(i)}\left( y_{0}^{N - 1},v_{0,0}^{i - 1} \middle| u_{2i} + u_{2i + 1} \right) \right. \nonumber \\
	&\left. \boxdot \boldsymbol{c}_{N}^{(i)}\left( y_{N}^{2N - 1},v_{1,0}^{i - 1} \middle| u_{2i + 1} \right) \right\rbrack
\end{align}
\begin{align} \label{eq:recoSp}
	&\boldsymbol{c}_{2N}^{(2i + 1)}\left( y_{0}^{2N - 1},u_{0}^{2i} \middle| u_{2i + 1} \right) \nonumber \\
	&= \boldsymbol{c}_{N}^{(i)}\left( y_{0}^{N - 1},v_{0,0}^{i - 1} \middle| u_{2i} + u_{2i + 1} \right) \boxdot \boldsymbol{c}_{N}^{(i)}\left( y_{N}^{2N - 1},v_{1,0}^{i - 1} \middle| u_{2i + 1} \right)
\end{align}
}
}

In summary, (\ref{eq:receSp})(\ref{eq:recoSp}) governs the recursion of log-density spectrum, which are driven by operators \(\boxdot\) and \(\boxplus\). Provided that (\ref{eq:receSp})(\ref{eq:recoSp}) admit closed-form representations, i.e.,
operators \(\boxdot\) and \(\boxplus\) could be expressed in
elementary functions, the recursion of log-density spectrum could be
implemented as an algorithm. According to (\ref{eq:def_odot})(\ref{eq:def_oplus}) and (\ref{eq:def_boxdot})(\ref{eq:def_boxplus}), the calculation of operators \(\boxdot\) and \(\boxplus\) involve integration, consequently, exact closed-form solutions of
(\ref{eq:receSp})(\ref{eq:recoSp}) are generally unavailable. Despite this, the following section develops
closed-form approximations of the recursions in (\ref{eq:receSp})(\ref{eq:recoSp}), leading to a
tractable algorithm of polar decoding under phase noise.

\section{Closed-Form Approximations of Log-Density Spectrum Recursions}

In order to implement polar decoding with a tractable algorithm, the
task of this section is to derive closed-form approximations of
log-density spectrum recursions described in (\ref{eq:receSp})(\ref{eq:recoSp}) of Section III.

\subsection{Approximation of the Phase Noise Model}

Firstly, for the convenience of spectral analysis, we apply an approximation to the phase noise model presented by (\ref{eq:PNden}), in order that its log-density spectrum consists of a finite number of spectral components. According to the Taylor expansion of the cosine function, there is a classical approximation that \(\cos{x} \approx 1-\frac{1}{2}x^2\), i.e., \(-\frac{1}{2}x^2 \approx \cos{x}-1\). Applying this approximation to (\ref{eq:PNden}), there is

{\small
	\begin{align}
		p_{\Phi_{n+1}|\Phi_{n}}\left( \phi_{n+1} \middle| \phi_{n} \right) \approx \frac{1}{\sqrt{2\pi}\sigma_{\phi}}\exp\left\{ \sigma_{\phi}^{- 2}\cos\left( \phi_{n + 1} - \phi_{n} \right) - \sigma_{\phi}^{2} \right\}
	\end{align}
}

More precisely, to ensure that this probability density integrates to one, a normalization is necessary, i.e.
{\small
\begin{align}
	p_{\Phi_{n+1}|\Phi_{n}}\left( \phi_{n + 1} \middle| \phi_{n} \right) \approx C_\phi^{-1} \exp\left\{\sigma_{\phi}^{- 2}\cos\left( \phi_{n + 1} - \phi_{n} \right)\right\}
\end{align}
}where
{\small\(C_\phi := \int_{0}^{2\pi}{\exp\left\{\sigma_{\phi}^{- 2}\cos\phi\right\}\mathrm{d}\phi}\)}.
Namely
{\small
\begin{align}
	p_{\Phi_{n+1}|\Phi_{n}}\left( \phi_{n + 1} \middle| \phi_{n} \right) \approx C_\phi^{-1} \exp\left\{ \mathfrak{R}\left\{ \sigma_{\phi}^{- 2}e^{j\left( \phi_{n + 1} - \phi_{n} \right)} \right\} \right\}
\end{align}
}

With this approximated phase noise model, it is ready to derive the log-density on the channel
\(L_{1}^{(0)}(y,\phi_{1}|x,\phi_{0})\):
{\footnotesize
\begin{align}
	&L_{1}^{(0)}\left( y,\phi_{1} \middle| x,\phi_{0} \right) = \ln\left( p_{\Phi_{1}|\Phi_{0}}\left( \phi_{1} \middle| \phi_{0} \right) p_{Y|X\Phi_0}\left( y \middle| x,\phi_{0} \right) \right) \nonumber \\
	&=\mathfrak{R}\left\{ \sigma_{\phi}^{- 2}e^{j\left( \phi_{1} - \phi_{0} \right)} \right\} - \frac{1}{2}\sigma^{-2}\left|y-e^{j\pi x}e^{j\phi_{0}} \right|^2 +C \nonumber \\
	&= \mathfrak{R}\left\{ \sigma_{\phi}^{- 2}e^{j\left( \phi_{0} - \phi_{1} \right)} + \sigma^{- 2}y^{*}e^{j\pi x}e^{j\phi_{0}} \right\} -\frac{1}{2}\sigma^{-2}\left(|y|^2+1\right) + C \nonumber \\
	&= \frac{1}{2}\sigma_{\phi}^{- 2}e^{j\phi_0}e^{-j\phi_1} + \frac{1}{2}\sigma_{\phi}^{- 2}e^{-j\phi_0}e^{j\phi_1} + \frac{1}{2}\sigma^{- 2}y^{*}e^{j\pi x}e^{j\phi_0}  \nonumber \\
	& + \frac{1}{2}\sigma^{-2}\left(y^{*}e^{j\pi x}\right)^*e^{-j\phi_0} - \frac{1}{2}\sigma^{-2}\left(|y|^2+1\right) + C
\end{align}
}where \(C:=-\ln{\left(2\pi\sigma^2 C_\phi\right)}\).

Therefore, its spectrum  \(\boldsymbol{c}_{1}^{(0)}\left( y \middle| x \right)\) is
{\footnotesize
\begin{align} \label{eq:chSp_ele}
	 &\boldsymbol{c}_{1}^{(0)}\left( y \middle| x \right) = \frac{1}{2}\sigma_{\phi}^{-2}\boldsymbol{\delta}_{1,-1} + \frac{1}{2}\sigma_{\phi}^{-2}\boldsymbol{\delta}_{-1,1}
	 +\frac{1}{2}\sigma^{- 2}y^{*}e^{j\pi x}\boldsymbol{\delta}_{1,0} \nonumber \\
	 &+ \frac{1}{2}\sigma^{-2}\left(y^{*}e^{j\pi x}\right)^*\boldsymbol{\delta}_{-1,0} + \left(-\frac{1}{2}\sigma^{-2}\left(|y|^2+1\right) + C\right)\boldsymbol{\delta}_{0,0}
\end{align}
}

Namely its spectrum \(\boldsymbol{c}_{1}^{(0)}\left( y \middle| x \right)\)
is supported on \(\overline{\mathcal{K}}\),
\(c_{1,\lbrack k,m\rbrack}^{(0)}\left( y \middle| x \right) = c_{1,\lbrack - k, - m\rbrack}^{(0)*}\left( y \middle| x \right)\),
and
{\footnotesize
\begin{align} \label{eq:chSp}
	\left\{ \begin{aligned}
	&c_{1,\lbrack 1, - 1\rbrack}^{(0)}\left( y \middle| x \right) = \frac{1}{2}\sigma_{\phi}^{- 2} \\
	&c_{1,\lbrack 1,0\rbrack}^{(0)}\left( y \middle| x \right) = \frac{1}{2}\sigma^{- 2}y^{*}e^{j\pi x} \\
	&c_{1,\lbrack 0,1\rbrack}^{(0)}\left( y \middle| x \right) = 0 \\
	&c_{1,\lbrack 0,0\rbrack}^{(0)}\left( y \middle| x \right) = -\frac{1}{2}\sigma^{-2}\left(|y|^2+1\right) + C
	\end{aligned} \right.
\end{align}
}

The support set
\(\overline{\mathcal{K}}\) is defined by
\begin{align} \label{eq:suppLast}
	\overline{\mathcal{K}} := \left\{ (k,m)||k + m| \leq 1,|k| \leq 1,|m| \leq 1 \right\}
\end{align}
i.e., {\small\(\overline{\mathcal{K}} = \left\{ (1,-1),(-1,1),(1,0),(-1,0),(0,1),(0,-1),(0,0)\right\}\)}.

According to (\ref{eq:chSp}), there is
\(c_{1,\lbrack 1,0\rbrack}^{(0)}\left( y \middle| x \right) = - c_{1,\lbrack 1,0\rbrack}^{(0)}\left( y \middle| \overline{x} \right)\),
\(c_{1,\lbrack 0,1\rbrack}^{(0)}\left( y \middle| x \right) = - c_{1,\lbrack 0,1\rbrack}^{(0)}\left( y \middle| \overline{x} \right)\),  \(c_{1,\lbrack 1, - 1\rbrack}^{(0)}\left( y \middle| x \right) = c_{1,\lbrack 1, - 1\rbrack}^{(0)}\left( y \middle| \overline{x} \right)\), \(c_{1,\lbrack 0, 0\rbrack}^{(0)}\left( y \middle| x \right) = c_{1,\lbrack 0,0\rbrack}^{(0)}\left( y \middle| \overline{x} \right)\).

With (\ref{eq:chSp}), it is ready to derive the approximated recursion formulas.

\subsection{Closed-Form Approximated Formulas of Recursions}

As presented in (\ref{eq:receSp})(\ref{eq:recoSp}), the recursions of log-density spectrum are driven by operators \(\boxdot\) and \(\boxplus\). Therefore, approximating recursions reduces to approximating operators \(\boxdot\) and \(\boxplus\). Approximating \(\boxdot\) and \(\boxplus\) for unrestricted inputs is challenging, however, the approximation of \(\boxdot\) and \(\boxplus\) becomes significantly simpler and more structured when the inputs are supported on \(\mathcal{K}\), which is defined by
\begin{align}
	\mathcal{K} := \left\{ (k,m) \middle| |k|+|m| \leq 2 \right\}
\end{align}
i.e., \(\widetilde{\mathcal{K}} = \left\{ (1,-1),(-1,1),(1,1),(-1,-1),(2,0),(-2,0),\right. \\
\left. (0,2),(0,-2),(1,0),(-1,0),(0,1),(0,-1),(0,0)\right\}\).

Namely, we consider
\(\boldsymbol{c}_{1} \boxdot \boldsymbol{c}_{2}\) and \(\boldsymbol{c}_{1} \boxplus \boldsymbol{c}_{2}\) when
\(\mathrm{supp}\left( \boldsymbol{c}_{1} \right),\mathrm{supp}\left( \boldsymbol{c}_{2} \right) \subseteq \mathcal{K}\). We discuss \(\boxdot\) and \(\boxplus\) respectively as follows:

1) With respect to \(\boxdot\), we obtain Proposition 1:

\emph{Proposition 1:} If
\(\mathrm{supp}\left( \boldsymbol{c}_{1} \right),\mathrm{supp}\left( \boldsymbol{c}_{2} \right) \subseteq \mathcal{K}\),
then there is
{\small
	\begin{align}
		&\boldsymbol{c}_{1} \boxdot \boldsymbol{c}_{2} = \boldsymbol{c}_1 \boldsymbol{\delta}_{-,0} + \boldsymbol{c}_2 \boldsymbol{\delta}_{0,-} + \left(h\left( \left| d \right|^{2} \right)+\ln{\pi}\right)\boldsymbol{\delta}_{0,0} \nonumber \\
		&+ \mathcal{F}\left\lbrack h\left( \mathcal{F}^{-1}\left[ \boldsymbol{c}_b * \boldsymbol{c}_b^{*-}\right] \right) \right\rbrack 
		+ \mathcal{F}\left\lbrack \overline{h}\left( \mathcal{F}^{-1}\left[ \boldsymbol{c}_b \right], d \right) \right\rbrack
	\end{align}
}where {\small\(\boldsymbol{c}_b = \left(\boldsymbol{c}_1 * \boldsymbol{\delta}_{-,-1}\right) \boldsymbol{\delta}_{-,0} + \left(\boldsymbol{c}_2 * \boldsymbol{\delta}_{-1,-}\right) \boldsymbol{\delta}_{0,-} \), \(d=c_{1,\lbrack 0,2\rbrack} + c_{2,\lbrack 2,0\rbrack}\),
and \(h(x) := \ln\left(\sum_{n = 0}^{\infty}{\frac{1}{{n!}^{2}}x^{n}}\right)\), \(\overline{h}\left( c_{1},c_{2} \right) := \ln\left( 1 + 2\sum_{k = 1}^{\infty}{{\overline{I}}_{2k}\left( \left| 2c_{1} \right| \right){\overline{I}}_{k}\left( \left| 2c_{2} \right| \right)\cos{k\left( \arg c_{2} - 2\arg c_{1} \right)}} \right)\),
and
\({\overline{I}}_{k}(x) := \frac{I_{k}(x)}{I_{0}(x)},x\in\mathbb{R}\).}.

The proof of Proposition 1 is in Appendix-B.

For approximation, we introduce Approximation 1, in order to omit the complicated
\(\overline{h}\) term.

\emph{Approximation 1:}
\(\ln\left( \int_{0}^{2\pi}{e^{\mathfrak{R}\left\{ c_{1}e^{j\phi} + c_{2}e^{j2\phi} \right\}}\mathrm{d}\phi} \right) \approx h\left( \left| \frac{1}{2}c_{1} \right|^{2} \right) + h\left( \left| \frac{1}{2}c_{2} \right|^{2} \right) + \ln{2\pi}\).

Furthermore, to simplify the \(h\) term, we apply a linearization to function \(h\), thereby leading to Approximation 2:

\emph{Approximation 2:} \(h(x) \approx \kappa x + \beta\), where
\(\kappa = \widehat{h^{'}}\left( x_{0} \right), \beta = \widehat{h}\left( x_{0} \right) - \widehat{h^{'}} \left( x_{0} \right)x_{0}\), and \(\widehat{h}(x_0) := \begin{cases}
	x_0 - \frac{1}{4}x_0^{2} + \frac{1}{9}x_0^{3}  &, x_0 \leq 2 \\
	2\sqrt{x_0} - \frac{1}{4}\ln x_0 + \frac{1}{16\sqrt{x_0}} - \frac{1}{2}\ln{4\pi}&, x_0 > 2
\end{cases}\), \(\widehat{h^{'}}(x_0):=\frac{1}{\frac{3}{16}+\sqrt{x_0+\frac{9}{16}}}\).

Here \(x_0\) is tuned to the specific case.

Applying Proposition 1 and Approximation 1,2, and tuning \(x_0\) to \(\left(\boldsymbol{c}_b * \boldsymbol{c}_b\right)_{[0,0]}\), there is
{\small
\begin{align} \label{eq:aprxDot_cmp}
	&\boldsymbol{c}_{1} \boxdot \boldsymbol{c}_{2} \approx \boldsymbol{c}_1 \boldsymbol{\delta}_{-,0} + \boldsymbol{c}_2 \boldsymbol{\delta}_{0,-} + \left(h\left( |d|^2 \right)+\beta+\ln{\pi}\right)\boldsymbol{\delta}_{0,0} \nonumber \\
	&+ \kappa\left(\boldsymbol{c}_b * \boldsymbol{c}_b^{*-}\right)
\end{align}
}where \(\kappa = \widehat{h^{'}}\left( D_1 \right),\beta = \widehat{h}\left( D_1 \right) - \widehat{h^{'}} \left( D_1 \right)D_1\), \(D_1=\left(\boldsymbol{c}_b * \boldsymbol{c}_b^{*-}\right)_{[0,0]}\).

Therefore, we define the approximated operator of \(\boxdot\), denoted by \(\widehat{\boxdot}\), as follows:

\textit{Definition 1:} Operator \(\widehat{\boxdot}\).
For \(\boldsymbol{c}_{1},\boldsymbol{c}_{2}\) satisfying \(\mathrm{supp}\left( \boldsymbol{c}_{1} \right),\mathrm{supp}\left( \boldsymbol{c}_{2} \right) \subseteq \mathcal{K}\), define \(\boldsymbol{c}_{1} \widehat{\boxdot} \boldsymbol{c}_{2}\) by
{\small
\begin{align} \label{eq:aprxdDot_cmp}
	&\boldsymbol{c}_{1} \widehat{\boxdot} \boldsymbol{c}_{2} := \boldsymbol{c}_1 \boldsymbol{\delta}_{-,0} + \boldsymbol{c}_2 \boldsymbol{\delta}_{0,-} + \left(\widehat{h}\left( D_2 \right)+\beta+\ln{\pi}\right)\boldsymbol{\delta}_{0,0} \nonumber \\
	&+ \kappa\left(\boldsymbol{c}_b * \boldsymbol{c}_b^{*-}\right)
\end{align}
}where
\(\boldsymbol{c}_b = \left(\boldsymbol{c}_1 * \boldsymbol{\delta}_{-,-1}\right) \boldsymbol{\delta}_{-,0} + \left(\boldsymbol{c}_2 * \boldsymbol{\delta}_{-1,-}\right) \boldsymbol{\delta}_{0,-} \),
\(\kappa = \widehat{h^{'}}\left( D_1 \right),\beta = \widehat{h}\left( D_1 \right) - \widehat{h^{'}} \left( D_1 \right)D_1\), \(D_1=\left(\boldsymbol{c}_b * \boldsymbol{c}_b^{*-}\right)_{[0,0]}\), \(D_2=\left| c_{1,\lbrack 0,2\rbrack} + c_{2,\lbrack 2,0\rbrack} \right|^{2}\).

2) With respect to \(\boxplus\), we obtain Proposition 2:

\textit{Proposition 2:} \(\boldsymbol{c}_1 \boxplus \boldsymbol{c}_2\mathord{=}\boldsymbol{c}_a \mathord{+} (\ln2)\boldsymbol{\delta}_{0,0} \mathord{+} \mathcal{F}\left[g\left(\mathcal{F}^{-1}\left[\boldsymbol{c}_d*\boldsymbol{c}_d\right]\right)\right]\), where \(\boldsymbol{c}_a = \frac{1}{2} (\boldsymbol{c}_1+\boldsymbol{c}_2)\), \(\boldsymbol{c}_d = \frac{1}{2} (\boldsymbol{c}_1-\boldsymbol{c}_2)\), and \(g(x) := \ln\left( \cosh\left( \sqrt{x} \right) \right),x \in \mathbb{R}^{+}\).

The proof of Proposition 2 is in Appendix-C.

The logarithmic hyperbolic cosine function
\(\ln\left( \cosh(x) \right)\) could be approximated by \(|x| - \ln 2\), i.e., \(g(x)\approx\sqrt{x}-\ln2\). Then a linearization could be applied to \(g(x)\):

\emph{Approximation 3:} \(g(x) \approx \frac{1}{2\sqrt{x_0}} x +\frac{1}{2}\sqrt{x_0} - \ln 2\).

Here \(x_0\) is tuned to the specific case.

Applying Proposition 2 and Approximation 2, and tuning \(x_0\) to \(\left(\boldsymbol{c}_d * \boldsymbol{c}_d\right)_{[0,0]}\), we have
\begin{align} \label{eq:aprxCmpPlus}
	\boldsymbol{c}_1 \boxplus \boldsymbol{c}_2 \approx \kappa \boldsymbol{c}_d*\boldsymbol{c}_d+\boldsymbol{c}_a + \beta \boldsymbol{\delta}_{0,0}
\end{align}
where \(\boldsymbol{c}_a = \frac{1}{2} (\boldsymbol{c}_1+\boldsymbol{c}_2), \boldsymbol{c}_d = \frac{1}{2} (\boldsymbol{c}_1-\boldsymbol{c}_2)\), and \(\kappa=\frac{1}{2\sqrt{D}}, \beta = \frac{1}{2}\sqrt{D}, D=\left(\boldsymbol{c}_d * \boldsymbol{c}_d\right)_{[0,0]}\).

To maintain the support set of \(\mathcal{K}\), we limit \(\boldsymbol{c}_d*\boldsymbol{c}_d\) to the support set of \(\mathcal{K}\), thereby presenting Approximation 4 as follows.

\textit{Approximation 4:} If \(\mathrm{supp}\left(\boldsymbol{c}\right) \subseteq \mathcal{K}\), then \(\boldsymbol{c}*\boldsymbol{c} \approx (\boldsymbol{c}*\boldsymbol{c}) \boldsymbol{\delta}_{\mathcal{K}}\), where \(\boldsymbol{\delta}_{\mathcal{K},[k,m]} :=\left\{ \begin{aligned}
	& 1,(k,m)\in\mathcal{K} \\
	& 0,\mathrm{otherwise}
\end{aligned} \right. \).

Further with Approximation 4, there is
\begin{align} \label{eq:aprxPlus_cmp1}
	\boldsymbol{c}_1 \boxplus \boldsymbol{c}_2 \approx \kappa \left(\boldsymbol{c}_d*\boldsymbol{c}_d\right)\boldsymbol{\delta}_{\mathcal{K}}+\boldsymbol{c}_a+\beta\boldsymbol{\delta}_{0,0}
\end{align}

Therefore, we define the approximated operator of \(\boxdot\), denoted by \(\widehat{\boxdot}\), as follows:

\textit{Definition 2:} Operator \(\widehat{\boxplus}\).
For \(\boldsymbol{c}_{1},\boldsymbol{c}_{2}\) satisfying \(\mathrm{supp}\left( \boldsymbol{c}_{1} \right),\mathrm{supp}\left( \boldsymbol{c}_{2} \right) \subseteq \mathcal{K}\), define \(\boldsymbol{c}_{1} \widehat{\boxplus} \boldsymbol{c}_{2}\) by
\begin{align} \label{eq:aprxdPlus_cmp}
	\boldsymbol{c}_1 \widehat{\boxplus} \boldsymbol{c}_2 = \kappa \left(\boldsymbol{c}_d*\boldsymbol{c}_d\right)\boldsymbol{\delta}_{\mathcal{K}}+\boldsymbol{c}_a+\beta\boldsymbol{\delta}_{0,0}
\end{align}
where \(\boldsymbol{c}_a = \frac{1}{2} (\boldsymbol{c}_1+\boldsymbol{c}_2)\), \(\boldsymbol{c}_d = \frac{1}{2} (\boldsymbol{c}_1-\boldsymbol{c}_2)\), and \(\kappa=\frac{1}{2\sqrt{D}}, \beta = \frac{1}{2}\sqrt{D}, D=\left(\boldsymbol{c}_d * \boldsymbol{c}_d\right)_{[0,0]}\).

In summary, when \(\mathrm{supp}\left( \boldsymbol{c}_{1} \right),\mathrm{supp}\left( \boldsymbol{c}_{2} \right) \subseteq \mathcal{K}\), operators \(\boxdot\) and \(\boxplus\) could be approximated by \(\widehat{\boxdot}\) and \(\widehat{\boxplus}\). According to the definition of \(\widehat{\boxdot}\) and \(\widehat{\boxplus}\), immediately there is a corollary, whose proof is omitted:

\textit{Corollary 1:} If
\(\mathrm{supp}\left( \boldsymbol{c}_{1} \right),\mathrm{supp}\left( \boldsymbol{c}_{2} \right) \subseteq \mathcal{K}\),
then \(\mathrm{supp}\left(\boldsymbol{c}_{1} \widehat{\boxdot} \boldsymbol{c}_{2}\right), \mathrm{supp}\left(\boldsymbol{c}_{1} \widehat{\boxplus} \boldsymbol{c}_{2}\right) \subseteq \mathcal{K}\).

Recalling that (\ref{eq:chSp_ele}) implies \(\mathrm{supp}\left(\boldsymbol{c}_{1}^{(0)}\left( y \middle| x \right)\right) \subseteq \overline{\mathcal{K}}\), and noting \(\overline{\mathcal{K}} \subseteq \mathcal{K}\), there is \(\mathrm{supp}\left(\boldsymbol{c}_{1}^{(0)}\left( y \middle| x \right)\right) \subseteq \mathcal{K}\). Therefore, the approximation by \(\widehat{\boxdot}\) and \(\widehat{\boxplus}\) is applicable to (\ref{eq:receSp})(\ref{eq:recoSp}) while \(N=1\), and according to Corollary 1, \(\widehat{\boldsymbol{c}}_{2}^{(0)}\left( y_{0}^{1} \middle| u_{0} \right), \widehat{\boldsymbol{c}}_{2}^{(1)}\left( y_{0}^{1},u_{0} \middle| u_{1} \right)\) are also supported on \(\mathcal{K}\). Recursively, we could apply
\(\widehat{\boxdot}\) and \(\widehat{\boxplus}\) to each
\(\boldsymbol{c}_{N}^{(i)}\). Namely
{\footnotesize
{\color{black}
\begin{align} \label{eq:aprxSRe_cmp}
	&\widehat{\boldsymbol{c}}_{2N}^{(2i)}\left( y_{0}^{2N - 1},u_{0}^{2i - 1} \middle| u_{2i} \right) = \nonumber \\
	&\underset{u_{2i + 1}\in\mathbb{ B}}{\widehat{\boxplus}}\left\lbrack \widehat{\boldsymbol{c}}_{N}^{(i)}\left( y_{0}^{N - 1},v_{0,0}^{i - 1} \middle| u_{2i} + u_{2i + 1} \right)\widehat{\boxdot}\widehat{\boldsymbol{c}}_{N}^{(i)}\left( y_{N}^{2N - 1},v_{1,0}^{i - 1} \middle| u_{2i + 1} \right) \right\rbrack
\end{align}
\begin{align} \label{eq:aprxSRo_cmp}
	&\widehat{\boldsymbol{c}}_{2N}^{(2i + 1)}\left( y_{0}^{2N - 1},u_{0}^{2i} \middle| u_{2i + 1} \right) = \nonumber \\
	& \widehat{\boldsymbol{c}}_{N}^{(i)}\left( y_{0}^{N - 1},v_{0,0}^{i - 1} \middle| u_{2i} + u_{2i + 1} \right)\widehat{\boxdot}\widehat{\boldsymbol{c}}_{N}^{(i)}\left( y_{N}^{2N - 1},v_{1,0}^{i - 1} \middle| u_{2i + 1} \right)
\end{align}
}
}

Driven by \(\widehat{\boxdot}\) and \(\widehat{\boxplus}\), (\ref{eq:aprxSRe_cmp})(\ref{eq:aprxSRo_cmp}) realize the general approximation of recursion formulas (\ref{eq:receSp})(\ref{eq:recoSp}).

\subsection{Simplifications of the Approximated Recursion Formulas}

Despite that \(\widehat{\boxdot}\) and \(\widehat{\boxplus}\) are generally applicable to approximating recursion formulas (\ref{eq:receSp})(\ref{eq:recoSp}), for the sake of computational efficiency, we specialize and tailor them according to different stages of the recursions. By specializing the general formulas according to specific stages, we can eliminate redundant computations and bypass unnecessary operations, thereby simplifying the approximated recursion formulas. Specifically, we notice that the support set could be further tightened according to the index \(i\), as presented in Lemma 1:

\textit{Proposition 3:} At each \(N\in\{2^n|n\in\mathbb{N}\}\), there is \(\mathrm{supp}\left( \widehat{\boldsymbol{c}}_{N}^{(N-1)} \left( y_{0}^{N - 1},u_{0}^{N-2} \middle| u_{N-1} \right) \right) \subseteq \overline{\mathcal{K}}\), \(\widehat{\boldsymbol{c}}_{N}^{(N-1)} \left( y_{0}^{N - 1},u_{0}^{N-2} \middle| 0 \right) \sim \widehat{\boldsymbol{c}}_{N}^{(N-1)} \left( y_{0}^{N - 1},u_{0}^{N-2} \middle| 1 \right)\); and \(\forall i<N-1\), \(\mathrm{supp}\left( \widehat{\boldsymbol{c}}_{N}^{(i)} \left( y_{0}^{N - 1},u_{0}^{i - 1} \middle| u_{i} \right) \right) \subseteq \widetilde{\mathcal{K}}\).

Here \(\boldsymbol{c}_1\sim\boldsymbol{c}_2\) is defined by \(\mathrm{supp}(\boldsymbol{c}_1-\boldsymbol{c}_2) \subseteq \{(k,m)||k|+|m|=1\}\), \(\mathrm{supp}(\boldsymbol{c}_1+\boldsymbol{c}_2) \subseteq \{(-1,1),(1,-1),(0,0)\}\), and \(\widetilde{\mathcal{K}}\) is defined
by
\begin{align} \label{eq:supp}
	\widetilde{\mathcal{K}} := \left\{ (k,m) \middle| |k| + |m| \in \left\{ 0,2 \right\} \right\}
\end{align}
i.e., \(\widetilde{\mathcal{K}} = \left\{ (1,-1),(-1,1),(1,1),(-1,-1),(2,0),(-2,0),\right. \\
\left. (0,2),(0,-2),(0,0)\right\}\).

The proof of Proposition 3 is in Appendix-D.

By leveraging these support sets tighter than \(\mathcal{K}\) according to \(i\), we can effectively bypass redundant calculations involving zero-valued entries. As a consequence, we simplify the computation of \(\widehat{\boxdot}\) and \(\widehat{\boxplus}\) under support set \(\overline{\mathcal{K}}\) and \(\widetilde{\mathcal{K}}\), as follows.

1) When \(\mathrm{supp}\left( \boldsymbol{c}_{1} \right),\mathrm{supp}\left( \boldsymbol{c}_{2} \right) \subseteq \overline{\mathcal{K}}\),  by expanding (\ref{eq:aprxdDot_cmp})(\ref{eq:aprxdPlus_cmp}) element-wise, \(\widehat{\boxdot}\) and \(\widehat{\boxplus}\) are simplified to \(\overline{\boxdot}\) and \(\overline{\boxplus}\), which are defined as follows:

{\color{black}
\emph{Definition 3:} Operator \(\overline{\boxdot}\).
For \(\boldsymbol{c}_{1},\boldsymbol{c}_{2}\) satisfying \(\mathrm{supp}\left( \boldsymbol{c}_{1} \right),\mathrm{supp}\left( \boldsymbol{c}_{2} \right) \subseteq \overline{\mathcal{K}}\), define
\(\boldsymbol{c}_{1}\overline{\boxdot}\boldsymbol{c}_{2} := \boldsymbol{c}_{s}\),
where \(\mathrm{supp}\left( \boldsymbol{c}_{s} \right) = \overline{\mathcal{K}}\),
\(c_{s,\lbrack k,m\rbrack} = c_{s,\lbrack - k, - m\rbrack}^{*}\), and
{\small
\begin{align} \label{eq:aprxLastDot}
	\left\{ \begin{aligned}
		& c_{s,\lbrack 1, - 1\rbrack} = \kappa c_{1,\lbrack 1, - 1\rbrack}c_{2,\lbrack 1, - 1\rbrack} \\
		& c_{s,\lbrack 1,0\rbrack} = \kappa c_{1,\lbrack 1, - 1\rbrack}\left( c_{1,\lbrack 0,1\rbrack} + c_{2,\lbrack 1,0\rbrack} \right) +c_{1,\lbrack 1,0\rbrack} \\
		& c_{s,\lbrack 0,1\rbrack} = \kappa c_{2,\lbrack 1, - 1\rbrack}^{*}\left( c_{1,\lbrack 0,1\rbrack} + c_{2,\lbrack 1,0\rbrack} \right) + c_{2,\lbrack 0,1\rbrack} \\
		& c_{s,\lbrack 0,0\rbrack} = \overline{h}\left( D \right) + c_{1,[0,0]}+c_{2,[0,0]} + \ln\pi
	\end{aligned} \right.
\end{align}
}where
\(\kappa = \widehat{h^{'}}\left(D\right)\), and \(D=\left| c_{1,\lbrack 1, - 1\rbrack} \right|^{2} + \left| c_{2,\lbrack 1, - 1\rbrack} \right|^{2} + \left| c_{1,\lbrack 0,1\rbrack} + c_{2,\lbrack 1,0\rbrack} \right|^{2}\).
}

{\color{black}
\emph{Definition 4:} Operator \(\overline{\boxplus}\).
For \(\boldsymbol{c}_{1},\boldsymbol{c}_{2}\) satisfying \(\mathrm{supp}\left( \boldsymbol{c}_{1} \right),\mathrm{supp}\left( \boldsymbol{c}_{2} \right) \subseteq \overline{\mathcal{K}}\), \(\boldsymbol{c}_1\sim\boldsymbol{c}_2\), define 
\(\boldsymbol{c}_{1}\overline{\boxplus}\boldsymbol{c}_{2} := \boldsymbol{c}_{s}\),
where \(\mathrm{supp}\left( \boldsymbol{c}_{s} \right) = \widetilde{\mathcal{K}}\),
\(c_{s,\lbrack k,m\rbrack} = c_{s,\lbrack - k, - m\rbrack}^{*}\), and
{\small
\begin{align} \label{eq:aprxLastPlus}
	\left\{ \begin{aligned}
		& c_{s,\lbrack 1,1\rbrack} = 2\kappa c_{1,\lbrack 1,0\rbrack}c_{1,\lbrack 0,1\rbrack} \\
		& c_{s,\lbrack 1, - 1\rbrack} = 2\kappa c_{1,\lbrack 1,0\rbrack}c_{1,\lbrack 0,1\rbrack}^{*} + c_{1,\lbrack 1, - 1\rbrack} \\
		& c_{s,\lbrack 2,0\rbrack} = \kappa c_{1,\lbrack 1,0\rbrack}^{2} \\
		& c_{s,\lbrack 0,2\rbrack} = \kappa c_{1,\lbrack 0,1\rbrack}^{2} \\
		& c_{s,\lbrack 0,0\rbrack} = \sqrt{D} + c_{1,\lbrack 0,0\rbrack}
	\end{aligned} \right.
\end{align}
}where \(\kappa  = \frac{1}{2\sqrt{D}}\), \(D=2\left( \left| c_{1,\lbrack 1,0\rbrack} \right|^{2} + \left| c_{1,\lbrack 0,1\rbrack} \right|^{2} \right)\).
}

2) When \(\mathrm{supp}\left( \boldsymbol{c}_{1} \right),\mathrm{supp}\left( \boldsymbol{c}_{2} \right) \subseteq \widetilde{\mathcal{K}}\), by expanding (\ref{eq:aprxdDot_cmp})(\ref{eq:aprxdPlus_cmp}) element-wise, \(\widehat{\boxdot}\) and \(\widehat{\boxplus}\) are simplified to \(\widetilde{\boxdot}\) and \(\widetilde{\boxplus}\), which are defined as follows:

{\color{black}
\emph{Definition 5:} Operator \(\widetilde{\boxdot}\).
For \(\boldsymbol{c}_{1},\boldsymbol{c}_{2}\) satisfying \(\mathrm{supp}\left( \boldsymbol{c}_{1} \right),\mathrm{supp}\left( \boldsymbol{c}_{2} \right) \subseteq \widetilde{\mathcal{K}}\), define \(\boldsymbol{c}_{1}\widetilde{\boxdot}\boldsymbol{c}_{2} := \boldsymbol{c}_{s}\),
where \(\mathrm{supp}\left( \boldsymbol{c}_{s} \right) = \widetilde{\mathcal{K}}\),
\(c_{s,\lbrack k,m\rbrack} = c_{s,\lbrack - k, - m\rbrack}^{*}\), and
{\small
\begin{align} \label{eq:aprxDot}
	\left\{ \begin{aligned}
		& c_{s,\lbrack 1, - 1\rbrack} = \kappa \left( c_{1,\lbrack 1,1\rbrack}c_{2,\lbrack 1,1\rbrack}^{*} + c_{1,\lbrack 1, - 1\rbrack}c_{2,\lbrack 1, - 1\rbrack} \right) \\
		& c_{s,\lbrack 1,1\rbrack} = \kappa \left( c_{1,\lbrack 1,1\rbrack}c_{2,\lbrack 1, - 1\rbrack}^{*} + c_{1,\lbrack 1, - 1\rbrack}c_{2,\lbrack 1,1\rbrack} \right) \\
		& c_{s,\lbrack 2,0\rbrack} = \kappa c_{1,\lbrack 1,1\rbrack}c_{1,\lbrack 1, - 1\rbrack} + c_{1,\lbrack 2,0\rbrack} \\
		& c_{s,\lbrack 0,2\rbrack} = \kappa c_{2,\lbrack 1,1\rbrack}c_{2,\lbrack 1, - 1\rbrack}^{*} + c_{2,\lbrack 0,2\rbrack} \\
		& c_{s,\lbrack 0,0\rbrack} = \widehat{h}\left( D_1 \right) + \widehat{h}\left( D_2 \right) + c_{1,[0,0]}+c_{2,[0,0]} + \ln\pi
	\end{aligned} \right.
\end{align}
where \(\kappa = \widehat{h^{'}}\left(D_1\right)\), and \(D_1=\left| c_{1,\lbrack 1, - 1\rbrack} \right|^{2} + \left| c_{1,\lbrack 1,1\rbrack} \right|^{2} + \left| c_{2,\lbrack 1, - 1\rbrack} \right|^{2} + \left| c_{2,\lbrack 1,1\rbrack} \right|^{2}\), \(D_2=\left| c_{1,\lbrack 0,2\rbrack} + c_{2,\lbrack 2,0\rbrack} \right|^{2}\).
}
}

{\color{black}
\emph{Definition 6:} Operator \(\widetilde{\boxplus}\).
For \(\boldsymbol{c}_{1},\boldsymbol{c}_{2}\) satisfying \(\mathrm{supp}\left( \boldsymbol{c}_{1} \right),\mathrm{supp}\left( \boldsymbol{c}_{2} \right) \subseteq \widetilde{\mathcal{K}}\), define \(\boldsymbol{c}_{1}\widetilde{\boxplus}\boldsymbol{c}_{2} := \boldsymbol{c}_{s}\),
where
\(\mathrm{supp}\left( \boldsymbol{c}_{s} \right) = \widetilde{\mathcal{K}}\),
\(c_{s,\lbrack k,m\rbrack} = c_{s,\lbrack - k, - m\rbrack}^{*}\), and
{\small
\begin{align} \label{eq:aprxPlus}
	\left\{ \begin{aligned}
		& c_{s,\lbrack 1,1\rbrack} = 2\kappa \left( c_{d,\lbrack 1, - 1\rbrack}^{*}c_{d,\lbrack 2,0\rbrack} + c_{d,\lbrack 1, - 1\rbrack}c_{d,\lbrack 0,2\rbrack} + c_{d,\lbrack 1,1\rbrack}c_{d,\lbrack 0,0\rbrack} \right) \\
		&+ c_{a,\lbrack 1,1\rbrack} \\
		& c_{s,\lbrack 1, - 1\rbrack} = 2\kappa \left( c_{d,\lbrack 1,1\rbrack}^{*}c_{d,\lbrack 2,0\rbrack} + c_{d,\lbrack 1,1\rbrack}c_{d,\lbrack 0,2\rbrack}^{*} + c_{d,\lbrack 1, - 1\rbrack}c_{d,\lbrack 0,0\rbrack} \right) \\
		&+ c_{a,\lbrack 1, - 1\rbrack} \\
		& c_{s,\lbrack 2,0\rbrack} = 2\kappa \left( c_{d,\lbrack 1,1\rbrack}c_{d,\lbrack 1, - 1\rbrack} + 2c_{d,\lbrack 2,0\rbrack}c_{d,\lbrack 0,0\rbrack} \right) + c_{a,\lbrack 2,0\rbrack} \\
		& c_{s,\lbrack 0,2\rbrack} = 2\kappa \left( c_{d,\lbrack 1,1\rbrack}c_{d,\lbrack 1, - 1\rbrack}^{*} + 2c_{d,\lbrack 0,2\rbrack}c_{d,\lbrack 0,0\rbrack} \right) + c_{a,\lbrack 0,2\rbrack} \\
		& c_{s,\lbrack 0,0\rbrack} = \sqrt{D} + c_{a,\lbrack 0,0\rbrack}
	\end{aligned} \right.
\end{align}
}where \(\boldsymbol{c}_a = \frac{1}{2} (\boldsymbol{c}_1+\boldsymbol{c}_2), \boldsymbol{c}_d = \frac{1}{2} (\boldsymbol{c}_1-\boldsymbol{c}_2)\), and \(\kappa  = \frac{1}{2\sqrt{D}}\), \(D=2\left( \left| c_{d,\lbrack 1, - 1\rbrack} \right|^{2} + \left| c_{d,\lbrack 1,1\rbrack} \right|^{2} + \left| c_{d,\lbrack 2,0\rbrack} \right|^{2} + \left| c_{d,\lbrack 0,2\rbrack} \right|^{2} \right)\).
}

In summary,
\(\widetilde{\boxdot}, \widetilde{\boxplus}\) are applicable to each
\(\boldsymbol{c}_{N}^{(i)},i < N - 1\), and
\(\overline{\boxdot}, \overline{\boxplus}\) are applicable to
\(\boldsymbol{c}_{N}^{(N - 1)}\). Namely
{\footnotesize
{\color{black}
\begin{align} \label{eq:aprxSRe}
	&\widehat{\boldsymbol{c}}_{2N}^{(2i)}\left( y_{0}^{2N - 1},u_{0}^{2i - 1} \middle| u_{2i} \right)= \nonumber \\
	& \left\{\begin{aligned}
		\underset{u_{2i + 1}\in \mathbb{B}}{\widetilde{\boxplus}}\left\lbrack \widehat{\boldsymbol{c}}_{N}^{(i)}\left( y_{0}^{N - 1},v_{0,0}^{i - 1} \middle| u_{2i} + u_{2i + 1} \right) \widetilde{\boxdot} \widehat{\boldsymbol{c}}_{N}^{(i)}\left( y_{N}^{2N - 1},v_{1,0}^{i - 1} \middle| u_{2i + 1} \right) \right\rbrack,\\i < N - 1 \\
		\underset{u_{2i + 1}\in\mathbb{ B}}{\overline{\boxplus}}\left\lbrack \widehat{\boldsymbol{c}}_{N}^{(i)}\left( y_{0}^{N - 1},v_{0,0}^{i - 1} \middle| u_{2i} + u_{2i + 1} \right)\overline{\boxdot}\widehat{\boldsymbol{c}}_{N}^{(i)}\left( y_{N}^{2N - 1},v_{1,0}^{i - 1} \middle| u_{2i + 1} \right) \right\rbrack,\\i = N - 1
	\end{aligned}\right.
\end{align}
\begin{align} \label{eq:aprxSRo}
	&\widehat{\boldsymbol{c}}_{2N}^{(2i + 1)}\left( y_{0}^{2N - 1},u_{0}^{2i} \middle| u_{2i + 1} \right)= \nonumber \\
	& \left\{\begin{aligned}
		\widehat{\boldsymbol{c}}_{N}^{(i)}\left( y_{0}^{N - 1},v_{0,0}^{i - 1} \middle| u_{2i} + u_{2i + 1} \right)\widetilde{\boxdot}\widehat{\boldsymbol{c}}_{N}^{(i)}\left( y_{N}^{2N - 1},v_{1,0}^{i - 1} \middle| u_{2i + 1} \right),\\i < N - 1 \\
		\widehat{\boldsymbol{c}}_{N}^{(i)}\left( y_{0}^{N - 1},v_{0,0}^{i - 1} \middle| u_{2i} + u_{2i + 1} \right)\overline{\boxdot}\widehat{\boldsymbol{c}}_{N}^{(i)}\left( y_{N}^{2N - 1},v_{1,0}^{i - 1} \middle| u_{2i + 1} \right),\\i = N - 1
	\end{aligned}\right.
\end{align}
}
}

By (\ref{eq:aprxSRe})(\ref{eq:aprxSRo}), the simplified approximated recursion formulas of log-density spectrum are
presented. The log-densities are represented by their
spectra, and the recursions among log-densities are carried out
approximately by closed-form computations among corresponding spectra.
Therefore, the proposed method is termed \emph{log-density spectrum recursion} (\emph{LDSR}).

Notably, the approximations adopted by LDSR, including Approximation 1-4, may introduce
approximation errors, thereby causing a certain degree of performance
degradation.

\section{Complexity Analysis}

As described in Section III-B, the ideal recursion formulas of log-density spectrum, i.e., (\ref{eq:receSp})(\ref{eq:recoSp}), generally do not admit closed-form representations, thereby not tractable. In stead, the approximated recursion formulas (\ref{eq:aprxSRe})(\ref{eq:aprxSRo}) are applied as a tractable decoding algorithm. Notably, for a floating-point number \(x\) stored by its mantissa \(m(x)\) and exponent \(a(x)\), i.e., \(x=m(x)\cdot2^{a(x)}, m(x)\in[1,2),a(x)\in\mathbb{Z}\), its logarithm \(\ln{x}\) could be approximated by \(\ln{x} \approx (\ln2)(m(x)+a(x)-0.943)\), and its square root \(\sqrt{x}\) could be approximated by \(m(\sqrt{x}) \approx \{\frac{1}{2}(m(x)+a(x)-0.943)\}+0.943, a(\sqrt{x}) \approx \lfloor \frac{1}{2}(m(x)+a(x)-0.943) \rfloor\). Therefore, the approximation of either \(\ln{x}\) or \(\sqrt{x}\) contains only 1 multiplication.
In the decoding algorithm, each \(\widetilde{\boxdot}\) operation involves 6 complex multiplications, 5
square norms and 1 \(\widehat{h}\), 1 \(\widehat{h^{'}}\) namely 3 multiplications, i.e., \(6 \times 4 + 5 \times 2 + 3 = 37\) real number
multiplications. Moreover, each \(\widetilde{\boxplus}\) operation involves 5 complex multiplications, 4 real-complex multiplications, 4
square norms and 1 square root, 1 reciprocal square root namely 2 multiplications, i.e., \(5 \times 4 + 4 \times 2 + 4 \times 2 + 2 = 38\) real number multiplications. Therefore, each non-last pair of bits has
\(37 \times 2 + 38 = 112\) real number multiplications.

For the last pair, the number of real number multiplications of \(\widehat{\boxdot}\) is \(3 \times 4 + 3 \times 2 + 3 = 21\), and of \(\widehat{\boxplus}\) is \(1 \times 4 + 2 \times 3 + 2 \times 2 + 2 = 16\). Therefore, each last pair of bits has \(21 \times 2 + 16 = 58\) real number multiplications.

The number of last pairs of all layers is \(\sum_{m=0}^{n-1}{2^m}=2^n-1=N-1\), so the number of non-last pairs is \(\frac{1}{2}N\log_{2}N-(N-1)\). Therefore, the number of real number multiplications is \(112 \times (\frac{1}{2}N\log_{2}N-(N-1)) + 58 \times (N-1) = 56N\log_{2}N - 54(N-1) = N(56\log_2{N}-54)+54\).

In summary, the
computational complexity is \(N(56\log_2{N}-54)+54\), in terms of real number
multiplication. Namely, the asymptotic complexity is still \(\mathcal{O}(N\log{N})\).

\section{Simulation Results}

\begin{figure*}[!t]
	\centering
	\subfloat[]{\includegraphics[width=2.3in]{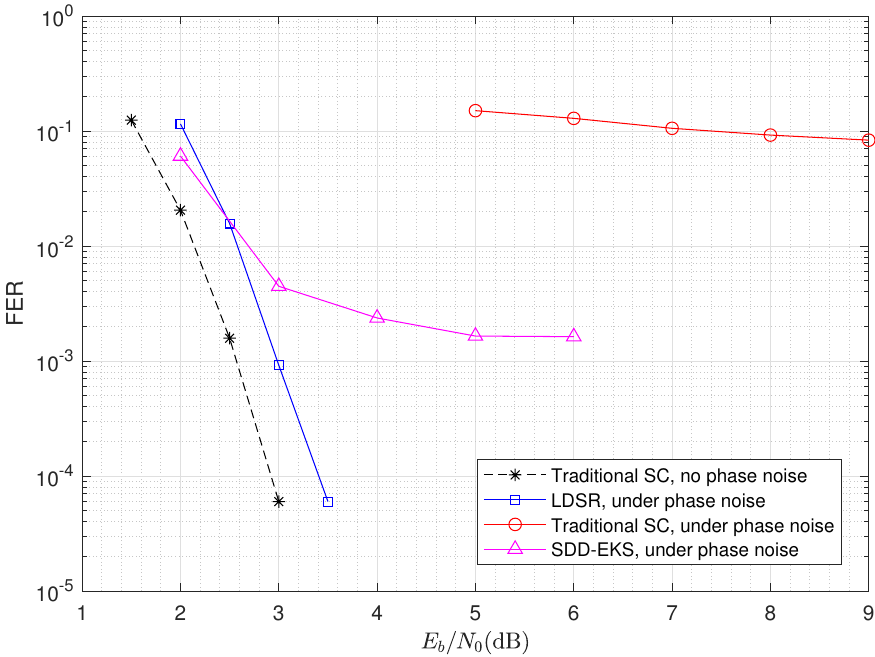}
		\label{a}}
	\hfil
	\subfloat[]{\includegraphics[width=2.3in]{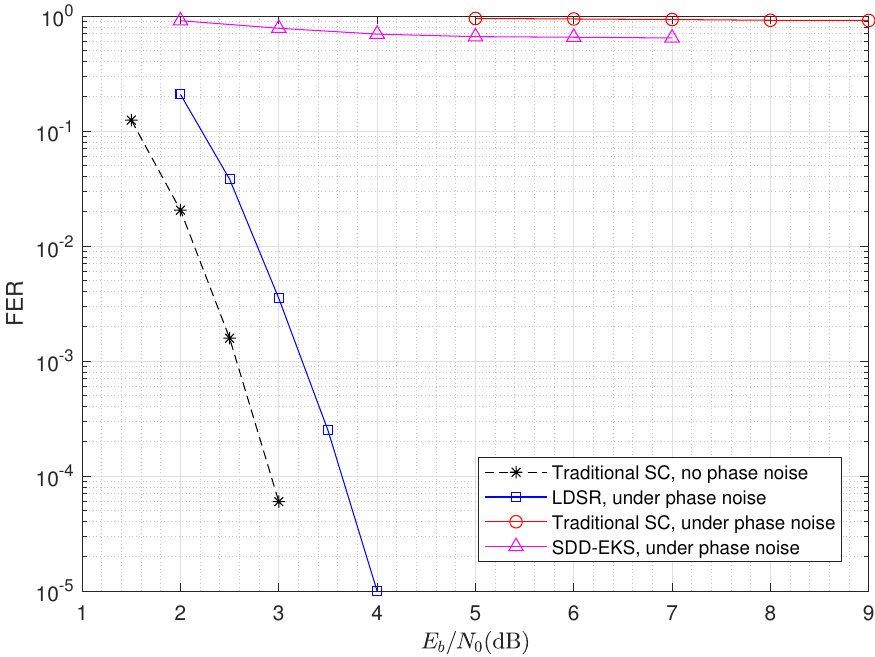}
		\label{b}}
	\hfil
	\subfloat[]{\includegraphics[width=2.3in]{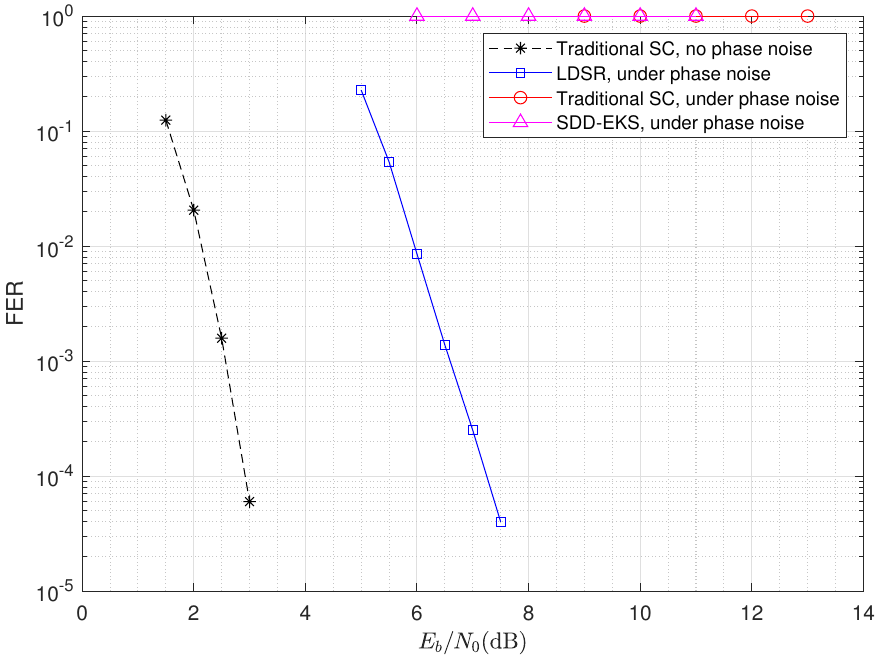}
		\label{c}}
	\caption{FERs under code length 512, code rate 1/2. (a)
		\(\sigma_{\phi} = \frac{1}{3}\frac{\pi}{32}\); (b)
		\(\sigma_{\phi} = \frac{1}{3}\frac{\pi}{8}\); (c)
		\(\sigma_{\phi} = \frac{1}{3}\frac{\pi}{2}\).}
	\label{fig:fers_512_256}
\end{figure*}

While the code length is 512, and the code rate is 1/2, under phase
noise, the frame error rates (FERs) are depicted in Fig. 3. The
modulation is BPSK. The phase noise model is the Wiener process
described in Section III-A. The phase noise transition standard
variation is
\(\sigma_{\phi} = \frac{1}{3}\frac{\pi}{32} = 0.0327 = 1.875{^\circ}\),
\(\sigma_{\phi} = \frac{1}{3}\frac{\pi}{8} = 0.1309 = 7.5{^\circ}\),
\(\sigma_{\phi} = \frac{1}{3}\frac{\pi}{2} = 0.5236 = 30{^\circ}\) in
Fig.~\ref{fig:fers_512_256}(a)-(c), respectively. As a Kalman-based benchmark, we implement
the soft decision directed extended Kalman smoother (SDD-EKS) proposed
in \cite{ref:Kalman Soft}. At each iteration, the EKS uses the decoder-generated soft
symbol estimates to estimate the phase noise. The received signal is
then phase-compensated using the smoothed estimate and passed back to
the decoder for the next iteration. The parameters required by the EKS
are the phase noise transition standard variation \(\sigma_{\phi}\) and
the AWGN standard variation \(\sigma\), and they are set to be known by
the EKS. As depicted in Fig.~\ref{fig:fers_512_256}, under strong phase noise, a significant
priority is achieved by the proposed LDSR decoding over traditional
polar decoding and Kalman-based decoding, and the error floor of
traditional SC and Kalman-based methods is solved by LDSR. Yet, under
weak phase noise, LDSR could be inferior to Kalman-based methods, which
could be attributed to the approximation errors of LDSR. Furthermore, by
comparing FERs under different \(\sigma_{\phi}\), evidently a remarkable
robustness of LDSR is illustrated.

\begin{figure*}[!t]
	\centering
	\subfloat[]{\includegraphics[width=1.7in]{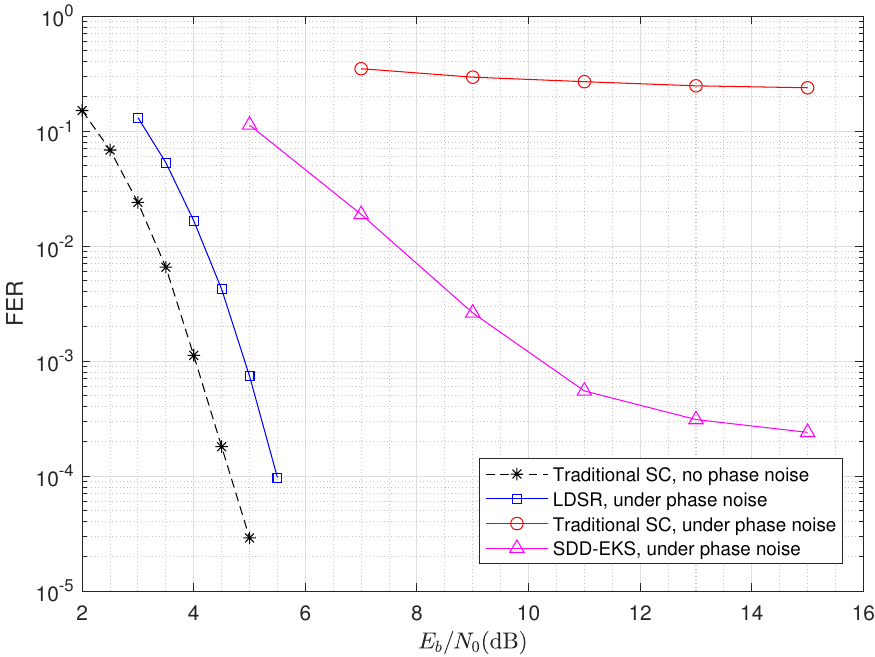}
		\label{a}}
	\hfil
	\subfloat[]{\includegraphics[width=1.7in]{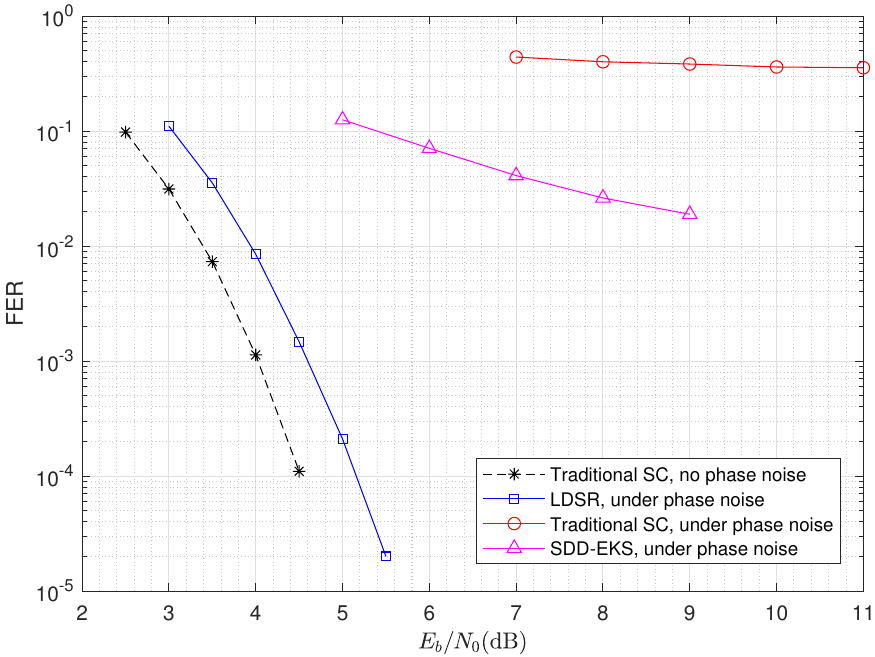}
		\label{b}}
	\hfil
	\subfloat[]{\includegraphics[width=1.7in]{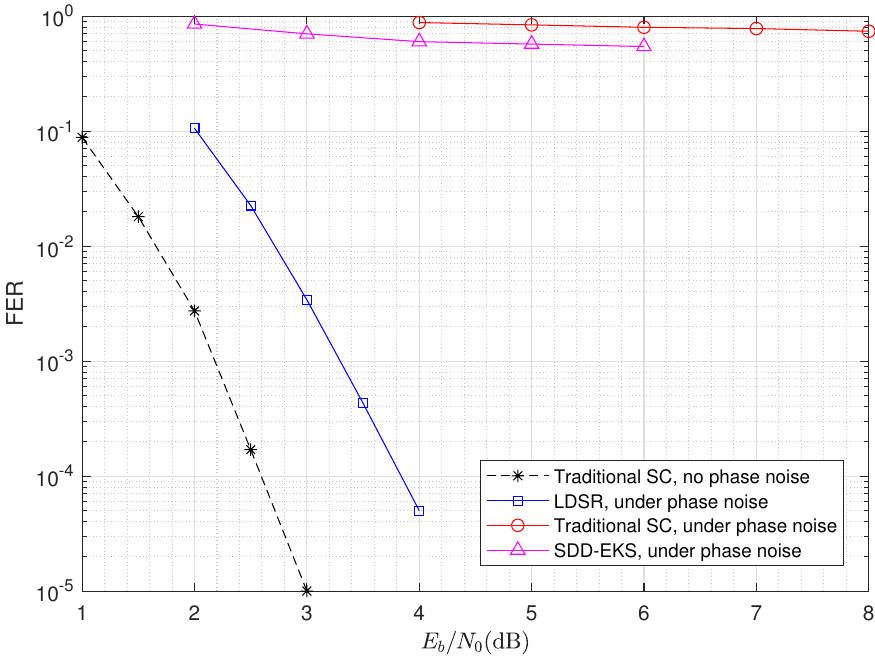}
		\label{c}}
	\hfil
	\subfloat[]{\includegraphics[width=1.7in]{media/fig_fers_512_256_sg8}
		\label{d}}
	\caption{FERs under
		\(\sigma_{\phi} = \frac{1}{3}\frac{\pi}{8}\), at different code lengths
		and rates. (a) At code length 128, rate 1/4; (b) At code length 128,
		rate 1/2; (c) At code length 512, rate 1/4; (d) At code length 512, rate
		1/2.}
	\label{fig:fersLenRts}
\end{figure*}

To observe the performance at different code lengths and rates, we
present the results under \(\sigma_{\phi} = \frac{1}{3}\frac{\pi}{8}\)
and at code length 128 and 512 code rate 1/4 and 1/2 in Fig.~\ref{fig:fersLenRts}. It is
shown in Fig.~\ref{fig:fersLenRts} that the superiority of proposed LDSR is remarkable
under various code lengths and rates, and this superiority becomes more
evident at larger code lengths and higher code rates.

\begin{figure*}[!t]
	\centering
	\subfloat[]{\includegraphics[width=2.3in]{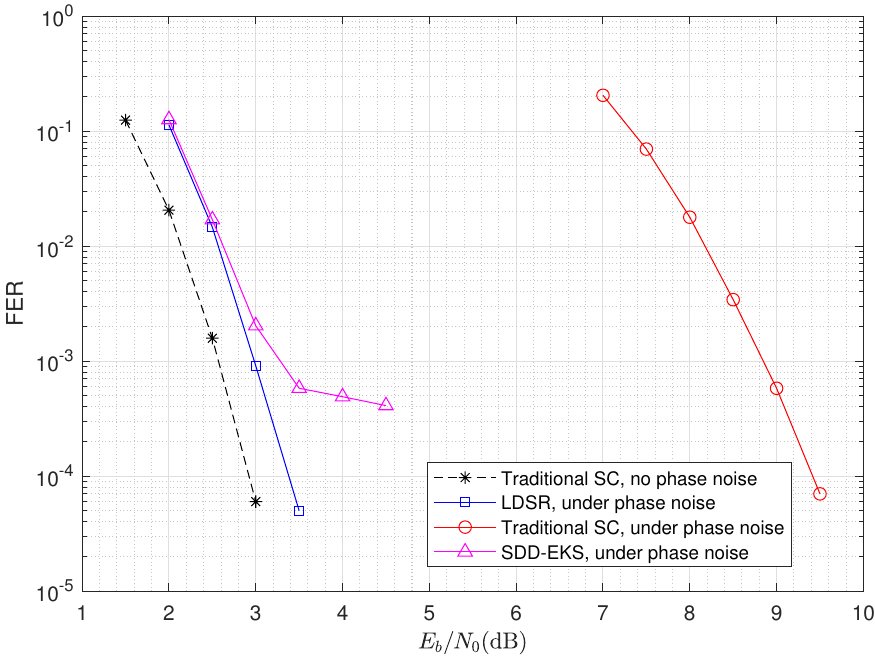}
		\label{a}}
	\hfil
	\subfloat[]{\includegraphics[width=2.3in]{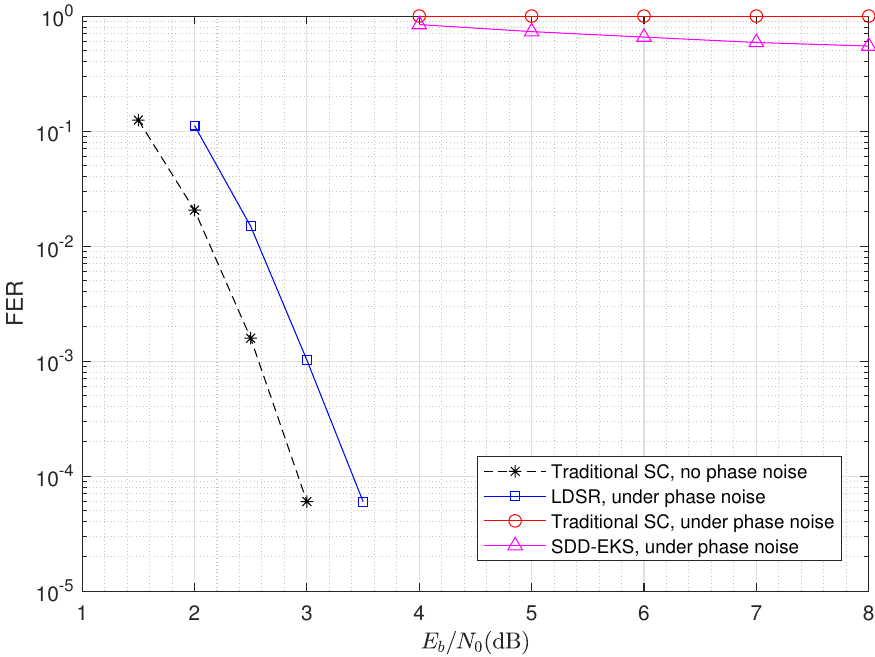}
		\label{b}}
	\hfil
	\subfloat[]{\includegraphics[width=2.3in]{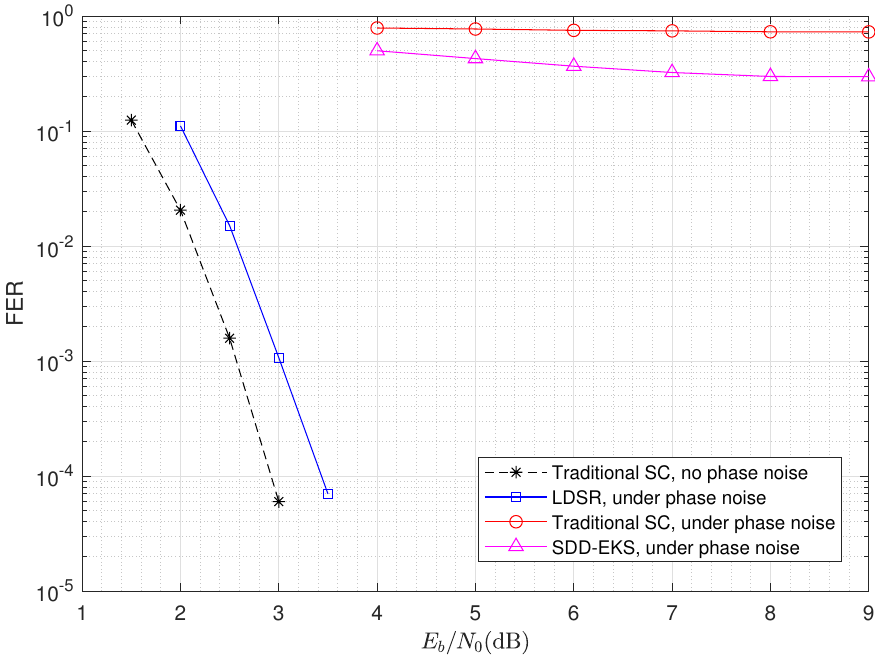}
		\label{c}}
	\caption{FERs at code length 512, code rate 1/2, under linear
		phase shift. (a) \(\Omega = \frac{\pi}{1024}\); (b)
		\(\Omega = \frac{\pi}{512}\); (c)
		\(\Omega \sim \mathcal{U}\left\lbrack 0,\frac{\pi}{256} \right\rbrack\).}
	\label{fig:fers_lin}
\end{figure*}

Moreover, to evaluate the generalizability of the proposed LDSR, namely,
whether LDSR remains effective beyond the original Wiener process model,
we further give FERs under a linear phase shift, where
\(\phi_{n} = \phi_{n - 1} + \Omega\), \(\Omega\) is the normalized
angular frequency, i.e., the phase increment per symbol. The code length
is 512 and code rate is 1/2. The normalized angular frequency \(\Omega\)
is \(\frac{\pi}{1024}\), \(\frac{\pi}{512}\), and uniformly distributed
over \(\left\lbrack 0,\frac{\pi}{256} \right\rbrack\) in Fig.~\ref{fig:fers_lin}(a)-(c),
respectively. The results under linear phase shift presented in Fig.~\ref{fig:fers_lin}
demonstrates the generalizability of LDSR.

\section{Conclusion}

A decoding method is proposed for polar codes under channels with strong
phase noise. In contrast to traditional SC decoding, where recursions
are performed on scalar soft information such as probabilities, the
proposed decoding recurses over probability density functions defined on
the phase-noise variable. To make the density-level recursion
computationally tractable, we represent each log-density by its
spectrum, and derive approximate formulas for the spectrum recursion,
therefore, the proposed method is termed log-density spectrum recursion
(LDSR). Numerical results are given under strong phase noise and AWGN,
showing that a prominent superiority of error performance is achieved by
LDSR in comparison with traditional SC and Kalman-based methods,
and that the error floor of traditional SC and Kalman-based methods is
solved by LDSR. Yet, under weak phase noise, LDSR could be inferior to
Kalman-based methods, which could be attributed to the approximation
errors of LDSR.

\appendix

\subsection{Proof of Equations (\ref{eq:rece})(\ref{eq:reco})}

There is
{\footnotesize
\begin{align*}
	&p_{Y_{0}^{2N - 1}U_{0}^{2i}\Phi_{2N}|U_{2i + 1}\Phi_{0}}\left( y_{0}^{2N - 1},u_{0}^{2i},\phi_{2N} \middle| u_{2i + 1},\ \phi_{0} \right) \\
	&= \frac{1}{2}\int_{0}^{2\pi}p_{Y_{0}^{2N - 1}U_{0}^{2i}\Phi_{2N}\Phi_{N}|U_{2i + 1}\Phi_{0}}\left( y_{0}^{2N - 1},u_{0}^{2i},\phi_{2N},\phi_{N} \middle| \right.\\
	&\left. u_{2i + 1},\phi_{0} \right)\mathrm{d}\phi_{N} \\
	&= \frac{1}{2}\int_{0}^{2\pi}p_{Y_{0}^{N - 1}V_{0,0}^{i - 1}\Phi_{N}|V_{0,i}\Phi_{0}}\left( y_{0}^{N - 1},v_{0,0}^{i - 1},\phi_{N} \middle| u_{2i} + u_{2i + 1},\phi_{0} \right) \\
	&\cdot p_{Y_{N}^{2N - 1}V_{1,0}^{i - 1}\Phi_{2N}|V_{1,i}\Phi_{N}Y_{0}^{N - 1}V_{0,0}^{i}\Phi_{0}}\left( y_{N}^{2N - 1},v_{1,0}^{i - 1},\phi_{2N} \middle| u_{2i + 1},\right.\\
	&\left.\phi_{N},y_{0}^{N - 1},v_{0,0}^{i},\phi_{0} \right)\mathrm{d}\phi_{N} \\
	&= \frac{1}{2}\int_{0}^{2\pi}p_{Y_{0}^{N - 1}V_{0,0}^{i - 1}\Phi_{N}|V_{0,i}\Phi_{0}}\left( y_{0}^{N - 1},v_{0,0}^{i - 1},\phi_{N} \middle| u_{2i} + u_{2i + 1},\phi_{0} \right) \\
	&\cdot p_{Y_{N}^{2N - 1}V_{1,0}^{i - 1}\Phi_{2N}|V_{1,i}\Phi_{N}}\left( y_{N}^{2N - 1},v_{1,0}^{i - 1},\phi_{2N} \middle| u_{2i + 1},\phi_{N} \right)\mathrm{d}\phi_{N}
\end{align*}
}

Therefore, (\ref{eq:reco}) holds.

Moreover, there is
{\footnotesize
\begin{align*}
	&p_{Y_{0}^{2N - 1}U_{0}^{2i - 1}\Phi_{2N}|U_{2i}\Phi_{0}}\left( y_{0}^{2N - 1},u_{0}^{2i - 1},\phi_{2N} \middle| u_{2i},\phi_{0} \right) \\
	&= \frac{1}{2}\sum_{u_{2i + 1}\in\mathbb{ B}}^{}{p_{Y_{0}^{2N - 1}U_{0}^{2i}\Phi_{2N}|U_{2i + 1}\Phi_{0}}\left( y_{0}^{2N - 1},u_{0}^{2i},\phi_{2N} \middle| u_{2i + 1},\ \phi_{0} \right)} \\
	&= \frac{1}{2}\sum_{u_{2i + 1}\in \mathbb{B}}^{}\int_{0}^{2\pi} p_{Y_{0}^{N - 1}V_{0,0}^{i - 1}\Phi_{N}|V_{0,i}\Phi_{0}}\left( y_{0}^{N - 1},v_{0,0}^{i - 1},\phi_{N} \middle| u_{2i} + u_{2i + 1},\right.\\
	&\left.\phi_{0} \right) 
	\cdot p_{Y_{N}^{2N - 1}V_{1,0}^{i - 1}\Phi_{2N}|V_{1,i}\Phi_{N}}\left( y_{N}^{2N - 1},v_{1,0}^{i - 1},\phi_{2N} \middle| u_{2i + 1},\phi_{N} \right)\mathrm{d}\phi_{N}
\end{align*}
}

Namely, (\ref{eq:rece}) holds. \(\hfill \blacksquare\)

\subsection{Proof of Proposition 1}

In preparation, we introduce Lemma 1:

\emph{Lemma 1:}
\(\ln\left( \int_{0}^{2\pi}{e^{\mathfrak{R}\left\{ c_{1}e^{j\phi} + c_{2}e^{j2\phi} \right\}}\mathrm{d}\phi} \right) = h\left( \left| \frac{1}{2}c_{1} \right|^{2} \right) + h\left( \left| \frac{1}{2}c_{2} \right|^{2} \right) + \ln{(2\pi)} + \overline{h}\left( \frac{1}{2}c_{1},\frac{1}{2}c_{2} \right)\),
where
\(\overline{h}\left( c_{1},c_{2} \right) := \ln\left( 1 + 2\sum_{k = 1}^{\infty}{{\overline{I}}_{2k}\left( \left| 2c_{1} \right| \right){\overline{I}}_{k}\left( \left| 2c_{2} \right| \right)\cos{k\left( \arg c_{2} - 2\arg c_{1} \right)}} \right)\),
and
\({\overline{I}}_{k}(x) := \frac{I_{k}(x)}{I_{0}(x)},x\in\mathbb{ R}\).

\textit{Proof of Lemma 1:} According to the Jacobi-Anger expression
\cite{ref:JA},
\(e^{\mathfrak{R}\left\{ z \right\}} = \sum_{k = - \infty}^{\infty}{I_{k}\left( |z| \right)e^{jk\arg z}}\),
then there is
{\small
	\begin{align*}
		&e^{\mathfrak{R}\left\{ c_{1}e^{j\phi} + c_{2}e^{j2\phi} \right\}} = e^{\mathfrak{R}\left\{ c_{1}e^{j\phi} \right\}} e^{\mathfrak{R}\left\{ c_{2}e^{j2\phi} \right\}} \\
		&= \sum_{n = - \infty}^{\infty}{I_{n}\left( \left| c_{1} \right| \right)e^{jn\left( \phi + \arg c_{1} \right)}}\sum_{m = - \infty}^{\infty}{I_{m}\left( \left| c_{2} \right| \right)e^{jm\left( 2\phi + \arg c_{2} \right)}} \\
		&= \sum_{l = - \infty}^{\infty}{e^{jl\phi}\sum_{k = - \infty}^{\infty}{I_{2k - l}\left( \left| c_{1} \right| \right)I_{k}\left( \left| c_{2} \right| \right)e^{jk\left( \arg c_{2} - 2\arg c_{1} \right)}}}
	\end{align*}
}
Note that while \(l \neq 0\),
\(\int_{0}^{2\pi}{e^{jl\phi}\mathrm{d}\phi} = 0\),
so
\(\int_{0}^{2\pi}{e^{\mathfrak{R}\left\{ c_{1}e^{j\phi} + c_{2}e^{j2\phi} \right\}}\mathrm{d}\phi} = 2\pi\sum_{k = - \infty}^{\infty}{I_{2k}\left( \left| c_{1} \right| \right)I_{k}\left( \left| c_{2} \right| \right)e^{jk\left( \arg c_{2} - 2\arg c_{1} \right)}}\),
i.e.
{\small
	\begin{align} \label{eq:biInt}
		&\int_{0}^{2\pi}{e^{\mathfrak{R}\left\{ c_{1}e^{j\phi} + c_{2}e^{j2\phi} \right\}}\mathrm{d}\phi} \nonumber \\
		&= 2\pi I_{0}\left( \left| c_{1} \right| \right)I_{0}\left( \left| c_{2} \right| \right) \nonumber \\
		&+ 4\pi\sum_{k = 1}^{\infty}{I_{2k}\left( \left| c_{1} \right| \right)I_{k}\left( \left| c_{2} \right| \right)\cos{k\left( \arg c_{2} - 2\arg c_{1} \right)}} \nonumber \\
		&= 2\pi I_{0}\left( \left| c_{1} \right| \right)I_{0}\left( \left| c_{2} \right| \right) \nonumber \\
		&\cdot \left\lbrack 1 + 2\sum_{k = 1}^{\infty}{{\overline{I}}_{2k}\left( \left| c_{1} \right| \right){\overline{I}}_{k}\left( \left| c_{2} \right| \right)\cos{k\left( \arg c_{2} - 2\arg c_{1} \right)}} \right\rbrack
	\end{align}
}the logarithm of which is
\(h\left( \left| \frac{1}{2}c_{1} \right|^{2} \right) + h\left( \left| \frac{1}{2}c_{2} \right|^{2} \right) + \ln{(2\pi)} + \overline{h}\left( \frac{1}{2}c_{1},\frac{1}{2}c_{2} \right)\). \(\hfill \blacksquare\)

With Lemma 2, it is sufficient to prove Proposition 1:

\textit{Proof of Proposition 1:} According to the definition of \(\boxdot\) and \(\odot\), there is
{\footnotesize
	\begin{align*}
		&\left( \mathcal{F}^{- 1}\left\lbrack \boldsymbol{c}_{1} \boxdot \boldsymbol{c}_{2} \right\rbrack \right)|_{\phi_{0},\phi_{2}} = \left( \mathcal{F}^{- 1}\left\lbrack \boldsymbol{c}_{1} \right\rbrack \odot \mathcal{F}^{- 1}\left\lbrack \boldsymbol{c}_{2} \right\rbrack \right)|_{\phi_{0},\phi_{2}} \\
		&= \ln\left( \frac{1}{2}\int_{0}^{2\pi}\exp\left\{ \sum_{k,m \in \mathcal{J}_2}^{}{c_{1,\lbrack k,m\rbrack}e^{jk\phi_{0}}e^{jm\phi_{1}}} \right.\right.\\
		&\left.\left.+ \sum_{p,q \in \mathcal{J}_2}^{}{c_{2,\lbrack p,q\rbrack}e^{jp\phi_{1}}e^{jq\phi_{2}}} \right\} \mathrm{d}\phi_{1} \right) \\
		&= \ln\left( \frac{1}{2}\int_{0}^{2\pi}\exp\left\{ \sum_{m \in \mathcal{J}_2}^{}e^{jm\phi_{1}} \sum_{k \in \mathcal{J}_2}^{}\left( c_{1,\lbrack k,m\rbrack}e^{jk\phi_{0}} \right.\right.\right.\\
		&\left.\left.\left.+ c_{2,\lbrack m,k\rbrack}e^{jk\phi_{2}} \right) \right\} \mathrm{d}\phi_{1} \right) \\
		&= \ln\left( \frac{1}{2}\int_{0}^{2\pi}\exp\left\{ \sum_{k \in \mathcal{J}_2}^{}\left( c_{1,\lbrack k,0\rbrack}e^{jk\phi_{0}} + c_{2,\lbrack 0,k\rbrack}e^{jk\phi_{2}} \right) \right.\right.\\
		&\left.\left.+ \mathfrak{R}\left\{ 2 e^{j\phi_{1}}\sum_{k \in \mathcal{J}_2}^{}{\left( c_{1,\lbrack k,1\rbrack}e^{jk\phi_{0}} + c_{2,\lbrack 1,k\rbrack}e^{jk\phi_{2}} \right)} \right\} \right.\right.\\
		&\left.\left.+ \mathfrak{R}\left\{2 e^{j2\phi_{1}} \left(c_{1,\lbrack 0,2\rbrack} + c_{2,\lbrack 2,0\rbrack}\right)\right\}\right\} \mathrm{d}\phi_{1} \right) \\
	\end{align*}
}where \(\mathcal{J}_2:=\{-2,-1,0,1,2\}\).

According to Lemma 2, there is
{\footnotesize
	\begin{align*}
		&\left( \mathcal{F}^{- 1}\left\lbrack \boldsymbol{c}_{1} \boxdot \boldsymbol{c}_{2} \right\rbrack \right)|_{\phi_{0},\phi_{2}}\\
		&= \sum_{k \in \mathcal{J}_2}^{}\left( c_{1,\lbrack k,0\rbrack}e^{jk\phi_{0}} + c_{2,\lbrack 0,k\rbrack}e^{jk\phi_{2}} \right) + \ln\pi \\
		&+ h\left( \left| \sum_{k \in \mathcal{J}_2}^{}{\left( c_{1,\lbrack k,1\rbrack}e^{jk\phi_{0}} + c_{2,\lbrack 1,k\rbrack}e^{jk\phi_{2}} \right)} \right|^{2} \right) \\
		&+ h\left( \left| c_{1,\lbrack 0,2\rbrack} + c_{2,\lbrack 2,0\rbrack} \right|^{2} \right) \\
		&+ \overline{h}\left( \sum_{k \in \mathcal{J}_2}^{}{\left( c_{1,\lbrack k,1\rbrack}e^{jk\phi_{0}} + c_{2,\lbrack 1,k\rbrack}e^{jk\phi_{2}} \right)}, c_{1,\lbrack 0,2\rbrack} + c_{2,\lbrack 2,0\rbrack} \right)
	\end{align*}
}

Namely, there is
{\small
	\begin{align*}
		&\mathcal{F}^{- 1}\left\lbrack \boldsymbol{c}_{1} \boxdot \boldsymbol{c}_{2} \right\rbrack = \mathcal{F}^{- 1}\left\lbrack \boldsymbol{c}_{1}\boldsymbol{\delta}_{- ,0} + \boldsymbol{c}_{2}\boldsymbol{\delta}_{0, -} \right\rbrack + h\left( \left| d \right|^{2} \right)+\ln{\pi} \\
		&+ h\left( \left| \mathcal{F}^{- 1}\left\lbrack \boldsymbol{c}_{b} \right\rbrack \right|^{2} \right) + \overline{h}\left( \mathcal{F}^{-1}\left[ \boldsymbol{c}_b \right], d \right)
	\end{align*}
}where
{\small\(\boldsymbol{c}_{b} = \left( \boldsymbol{c}_{1}*\boldsymbol{\delta}_{- , - 1} \right)\boldsymbol{\delta}_{- ,0} + \left( \boldsymbol{c}_{2}*\boldsymbol{\delta}_{- 1, -} \right)\boldsymbol{\delta}_{0, -}\), \(d=c_{1,\lbrack 0,2\rbrack} + c_{2,\lbrack 2,0\rbrack}\)}.

Then there is
{\small
	\begin{align*}
		&\boldsymbol{c}_{1} \boxdot \boldsymbol{c}_{2} = \boldsymbol{c}_{1}\boldsymbol{\delta}_{- ,0} + \boldsymbol{c}_{2}\boldsymbol{\delta}_{0, -} + \left(h\left( \left| d \right|^{2} \right)+\ln{\pi}\right)\boldsymbol{\delta}_{0,0}\\
		& + \mathcal{F}\left\lbrack h\left( \left| \mathcal{F}^{- 1}\left\lbrack \boldsymbol{c}_{b} \right\rbrack \right|^{2} \right) \right\rbrack + \mathcal{F}\left[\overline{h}\left( \mathcal{F}^{-1}\left[ \boldsymbol{c}_b \right], d \right)\right] \\
		&= \boldsymbol{c}_1 \boldsymbol{\delta}_{-,0} + \boldsymbol{c}_2 \boldsymbol{\delta}_{0,-} + \left(h\left( \left| d \right|^{2} \right)+\ln{\pi}\right)\boldsymbol{\delta}_{0,0} \\
		&+ \mathcal{F}\left\lbrack h\left( \mathcal{F}^{-1}\left[ \boldsymbol{c}_b * \boldsymbol{c}_b^{*-}\right] \right) \right\rbrack 
		+ \mathcal{F}\left\lbrack \overline{h}\left( \mathcal{F}^{-1}\left[ \boldsymbol{c}_b \right], d \right) \right\rbrack
	\end{align*}
} \(\hfill \blacksquare\)

\subsection{Proof of Proposition 2}

According to the definition of \(\boxplus\), there is
{\small
	\begin{align*}
		&\mathcal{F}^{-1}\left[\boldsymbol{c}_1 \boxplus \boldsymbol{c}_2\right]=\ln\left( \exp\left\{\mathcal{F}^{-1}[\boldsymbol{c}_1]\right\} + \exp\left\{\mathcal{F}^{-1}[\boldsymbol{c}_2]\right\}\right)\\
		&=\frac{1}{2} \left(\mathcal{F}^{-1}[\boldsymbol{c}_1]+\mathcal{F}^{-1}[\boldsymbol{c}_2]\right) \\
		&+ \ln\left(2\cosh\left(\frac{1}{2} \left(\mathcal{F}^{-1}[\boldsymbol{c}_1]-\mathcal{F}^{-1}[\boldsymbol{c}_2]\right)\right)\right) \\
		&=\mathcal{F}^{-1}\left[\frac{1}{2}\left(\boldsymbol{c}_1+\boldsymbol{c}_2\right)\right] + \ln\left(2\cosh\left(\mathcal{F}^{-1}\left[\frac{1}{2}\left(\boldsymbol{c}_1-\boldsymbol{c}_2\right)\right]\right)\right) \\
	\end{align*}
}

Denoting \(\boldsymbol{c}_a = \frac{1}{2} (\boldsymbol{c}_1+\boldsymbol{c}_2), \boldsymbol{c}_d = \frac{1}{2} (\boldsymbol{c}_1-\boldsymbol{c}_2)\), there is
{\small
	\begin{align*}
		&\mathcal{F}^{-1}\left[\boldsymbol{c}_1 \boxplus \boldsymbol{c}_2\right]=\mathcal{F}^{-1}\left[\boldsymbol{c}_a\right] + \ln\left(2\cosh\left(\mathcal{F}^{-1}\left[\boldsymbol{c}_d\right]\right)\right) \\
		&=\mathcal{F}^{-1}\left[\boldsymbol{c}_a\right] +\ln2+ \ln\left(\cosh\left(\mathcal{F}^{-1}\left[\boldsymbol{c}_d\right]\right)\right) \\
		&=\mathcal{F}^{-1}\left[\boldsymbol{c}_a\right] +\ln2+ g\left(\left(\mathcal{F}^{-1}\left[\boldsymbol{c}_d\right]\right)^2\right) \\
		&=\mathcal{F}^{-1}\left[\boldsymbol{c}_a\right] +\ln2+ g\left(\mathcal{F}^{-1}\left[\boldsymbol{c}_d*\boldsymbol{c}_d\right]\right)
	\end{align*}
}

Therefore, there is
{\small
	\begin{align*}
		\boldsymbol{c}_1 \boxplus \boldsymbol{c}_2=\boldsymbol{c}_a + (\ln2)\boldsymbol{\delta}_{0,0} + \mathcal{F}\left[g\left(\mathcal{F}^{-1}\left[\boldsymbol{c}_d*\boldsymbol{c}_d\right]\right)\right]
	\end{align*}
} \(\hfill \blacksquare\)

\subsection{Proof of Proposition 3}

In preparation, we introduce Lemma 2, 3:

\textit{Lemma 2:} If
\(\mathrm{supp}\left( \boldsymbol{c}_{1} \right),\mathrm{supp}\left( \boldsymbol{c}_{2} \right) \subseteq \overline{\mathcal{K}}\),  \(\boldsymbol{c}_1\sim\boldsymbol{c}_2\), then \(\mathrm{supp}\left(\boldsymbol{c}_{1} \widehat{\boxdot} \boldsymbol{c}_{2}\right) \subseteq \overline{\mathcal{K}}\), \(\mathrm{supp}\left(\boldsymbol{c}_{1} \widehat{\boxplus} \boldsymbol{c}_{2}\right) \subseteq \widetilde{\mathcal{K}}\); and if \(\mathrm{supp}\left( \boldsymbol{c}_{1} \right),\\ \mathrm{supp}\left( \boldsymbol{c}_{2} \right) \subseteq \widetilde{\mathcal{K}}\), then \(\mathrm{supp}\left(\boldsymbol{c}_{1} \widehat{\boxdot} \boldsymbol{c}_{2}\right), \mathrm{supp}\left(\boldsymbol{c}_{1} \widehat{\boxplus} \boldsymbol{c}_{2}\right) \subseteq \widetilde{\mathcal{K}}\).

\textit{Proof of Lemma 2:} If \(\mathrm{supp}\left( \boldsymbol{c}_{1} \right),\mathrm{supp}\left( \boldsymbol{c}_{2} \right) \subseteq \overline{\mathcal{K}}\), \(\boldsymbol{c}_1\sim\boldsymbol{c}_2\), then in (\ref{eq:aprxdDot_cmp}), \(\mathrm{supp}(\boldsymbol{c}_b) \subseteq \{(-1,0),(0,-1),(0,0)\}\), so \(\mathrm{supp}\left(\boldsymbol{c}_b * \boldsymbol{c}_b^{*-}\right) \subseteq \overline{\mathcal{K}}\), and \(\mathrm{supp}\left(\boldsymbol{c}_1 \boldsymbol{\delta}_{-,0}\right)\subseteq\{(k,m)|m=0,|k|\leq1\}\subseteq \overline{\mathcal{K}}\), \(\mathrm{supp}\left(\boldsymbol{c}_1 \boldsymbol{\delta}_{0,-}\right)\subseteq\{(k,m)|k=0,|m|\leq1\}\subseteq \overline{\mathcal{K}}\), \(\mathrm{supp}\left(\boldsymbol{\delta}_{0,0}\right)\subseteq \overline{\mathcal{K}}\), therefore, \(\mathrm{supp}\left(\boldsymbol{c}_{1} \widehat{\boxdot} \boldsymbol{c}_{2}\right)\subseteq \overline{\mathcal{K}}\); and in (\ref{eq:aprxdPlus_cmp}), there is \(\mathrm{supp}(\boldsymbol{c}_d)=\mathrm{supp}(\boldsymbol{c}_1-\boldsymbol{c}_2) \subseteq \{(k,m)||k|+|m|=1\}\), so \(\mathrm{supp}\left(\left(\boldsymbol{c}_d*\boldsymbol{c}_d\right)\boldsymbol{\delta}_{\mathcal{K}}\right) \subseteq \widetilde{\mathcal{K}}\), and \(\mathrm{supp}\left(\boldsymbol{c}_a\right) = \mathrm{supp}(\boldsymbol{c}_1+\boldsymbol{c}_2) \subseteq \{(-1,1),(1,-1),(0,0)\} \subseteq \widetilde{\mathcal{K}}\), \(\mathrm{supp}\left(\boldsymbol{\delta}_{0,0}\right)\subseteq \widetilde{\mathcal{K}}\), therefore, \(\mathrm{supp}\left(\boldsymbol{c}_{1} \widehat{\boxplus} \boldsymbol{c}_{2}\right) \subseteq \widetilde{\mathcal{K}}\).

If \(\mathrm{supp}\left( \boldsymbol{c}_{1} \right),\mathrm{supp}\left( \boldsymbol{c}_{2} \right) \subseteq \widetilde{\mathcal{K}}\), then in (\ref{eq:aprxdDot_cmp}), \(\mathrm{supp}(\boldsymbol{c}_b) \subseteq \{(k,m)||k|+|m|=1\}\), so \(\mathrm{supp}\left(\boldsymbol{c}_b * \boldsymbol{c}_b^{*-}\right) \subseteq \widetilde{\mathcal{K}}\), and \(\mathrm{supp}\left(\boldsymbol{c}_1 \boldsymbol{\delta}_{-,0}\right)\subseteq\{(k,m)|m=0,|k|=0,2\}\subseteq \widetilde{\mathcal{K}}\), \(\mathrm{supp}\left(\boldsymbol{c}_1 \boldsymbol{\delta}_{0,-}\right)\subseteq\{(k,m)|k=0,|m|=0,2\}\subseteq \widetilde{\mathcal{K}}\), \(\mathrm{supp}\left(\boldsymbol{\delta}_{0,0}\right)\subseteq \widetilde{\mathcal{K}}\), therefore, \(\mathrm{supp}\left(\boldsymbol{c}_{1} \widehat{\boxdot} \boldsymbol{c}_{2}\right)\subseteq \widetilde{\mathcal{K}}\); and in (\ref{eq:aprxdPlus_cmp}), \(\mathrm{supp}(\boldsymbol{c}_d) \subseteq \widetilde{\mathcal{K}}\), so \(\mathrm{supp}\left(\left(\boldsymbol{c}_d*\boldsymbol{c}_d\right)\boldsymbol{\delta}_{\mathcal{K}}\right) \subseteq \widetilde{\mathcal{K}}\), and \(\mathrm{supp}\left(\boldsymbol{c}_a\right) \subseteq \widetilde{\mathcal{K}}\), \(\mathrm{supp}\left(\boldsymbol{\delta}_{0,0}\right)\subseteq \widetilde{\mathcal{K}}\), therefore, \(\mathrm{supp}\left(\boldsymbol{c}_{1} \widehat{\boxplus} \boldsymbol{c}_{2}\right) \subseteq \widetilde{\mathcal{K}}\).
\(\hfill \blacksquare\)

\textit{Lemma 3:} If \(\mathrm{supp}(\boldsymbol{c}_{1}),\mathrm{supp}(\boldsymbol{c}_{2}),\mathrm{supp}(\boldsymbol{c}_{1}^{\prime}),\mathrm{supp}(\boldsymbol{c}_{2}^{\prime})\subseteq\overline{\mathcal{K}}\) and \(\boldsymbol{c}_{1} \sim \boldsymbol{c}_{1}^{\prime}\), \(\boldsymbol{c}_{2} \sim \boldsymbol{c}_{2}^{\prime}\), then \(\boldsymbol{c}_1\widehat{\boxdot}\boldsymbol{c}_2 \sim \boldsymbol{c}_1^{\prime}\widehat{\boxdot}\boldsymbol{c}_2^{\prime}\).

\textit{Proof of Lemma 3:} According to (\ref{eq:aprxdDot_cmp}), when \(\mathrm{supp}(\boldsymbol{c}_{1}),\mathrm{supp}(\boldsymbol{c}_{2})\subseteq\overline{\mathcal{K}}\), \(\widehat{\boxdot}\) is simplified to:
{\small
\begin{align} \label{eq:aprxdDot_cmp_simp}
	\boldsymbol{c}_{1} \widehat{\boxdot} \boldsymbol{c}_{2} := \boldsymbol{c}_1 \boldsymbol{\delta}_{-,0} + \boldsymbol{c}_2 \boldsymbol{\delta}_{0,-} + \left(\beta+\ln{\pi}\right)\boldsymbol{\delta}_{0,0} + \kappa\left(\boldsymbol{c}_b * \boldsymbol{c}_b^{*-}\right)
\end{align}
}

Denote \(\boldsymbol{c}_{1d}:=\frac{1}{2}(\boldsymbol{c}_{1} - \boldsymbol{c}_{1}^{\prime})\), \(\boldsymbol{c}_{1a}:=\frac{1}{2}(\boldsymbol{c}_{1} + \boldsymbol{c}_{1}^{\prime})\), \(\boldsymbol{c}_{2d}:=\frac{1}{2}(\boldsymbol{c}_{2} - \boldsymbol{c}_{2}^{\prime})\), \(\boldsymbol{c}_{2a}:=\frac{1}{2}(\boldsymbol{c}_{2} + \boldsymbol{c}_{2}^{\prime})\), then \(\boldsymbol{c}_{1}=\boldsymbol{c}_{1a}+\boldsymbol{c}_{1d}\), \(\boldsymbol{c}_{1}^{\prime}=\boldsymbol{c}_{1a}-\boldsymbol{c}_{1d}\), and \(\boldsymbol{c}_{2}=\boldsymbol{c}_{2a}+\boldsymbol{c}_{2d}\), \(\boldsymbol{c}_{2}^{\prime}=\boldsymbol{c}_{2a}-\boldsymbol{c}_{2d}\). Substituting \(\boldsymbol{c}_{1}=\boldsymbol{c}_{1a}+\boldsymbol{c}_{1d}\), \(\boldsymbol{c}_{2}=\boldsymbol{c}_{2a}+\boldsymbol{c}_{2d}\) into (\ref{eq:aprxdDot_cmp_simp}) and simplifying, we obtain
{\small
\begin{align*}
	\boldsymbol{c}_{1} \widehat{\boxdot} \boldsymbol{c}_{2}
	&=\boldsymbol{c}_{1a} \boldsymbol{\delta}_{-,0} + \boldsymbol{c}_{2a} \boldsymbol{\delta}_{0,-} +\left(\beta+\ln{\pi}\right)\boldsymbol{\delta}_{0,0} \\
	&+ \kappa\left(\boldsymbol{c}_{ba}*\boldsymbol{c}_{ba}^{*-}+\boldsymbol{c}_{bd}*\boldsymbol{c}_{bd}^{*-}\right) \\
	&+ \boldsymbol{c}_{1d} \boldsymbol{\delta}_{-,0} + \boldsymbol{c}_{2d} \boldsymbol{\delta}_{0,-} + \kappa\left(\boldsymbol{c}_{ba}*\boldsymbol{c}_{bd}^{*-}+\boldsymbol{c}_{bd}*\boldsymbol{c}_{ba}^{*-}\right)
\end{align*}
}and substituting \(\boldsymbol{c}_{1}^{\prime}=\boldsymbol{c}_{1a}-\boldsymbol{c}_{1d}\), \(\boldsymbol{c}_{2}^{\prime}=\boldsymbol{c}_{2a}-\boldsymbol{c}_{2d}\) into (\ref{eq:aprxdDot_cmp_simp}) and simplifying, we obtain
{\small
\begin{align*}
	\boldsymbol{c}_{1}^{\prime} \widehat{\boxdot} \boldsymbol{c}_{2}^{\prime}
	&=\boldsymbol{c}_{1a} \boldsymbol{\delta}_{-,0} + \boldsymbol{c}_{2a} \boldsymbol{\delta}_{0,-} +\left(\beta+\ln{\pi}\right)\boldsymbol{\delta}_{0,0} \\
	&+ \kappa\left(\boldsymbol{c}_{ba}*\boldsymbol{c}_{ba}^{*-}+\boldsymbol{c}_{bd}*\boldsymbol{c}_{bd}^{*-}\right) \\
	&- \boldsymbol{c}_{1d} \boldsymbol{\delta}_{-,0} - \boldsymbol{c}_{2d} \boldsymbol{\delta}_{0,-} - \kappa\left(\boldsymbol{c}_{ba}*\boldsymbol{c}_{bd}^{*-}+\boldsymbol{c}_{bd}*\boldsymbol{c}_{ba}^{*-}\right)
\end{align*}
}where \(\boldsymbol{c}_{ba}=\left(\boldsymbol{c}_{1a} * \boldsymbol{\delta}_{-,-1}\right) \boldsymbol{\delta}_{-,0} + \left(\boldsymbol{c}_{2a} * \boldsymbol{\delta}_{-1,-}\right) \boldsymbol{\delta}_{0,-}\) and \(\boldsymbol{c}_{bd}=\left(\boldsymbol{c}_{1d} * \boldsymbol{\delta}_{-,-1}\right) \boldsymbol{\delta}_{-,0} + \left(\boldsymbol{c}_{2d} * \boldsymbol{\delta}_{-1,-}\right) \boldsymbol{\delta}_{0,-}\).

Therefore, \(\frac{1}{2}(\boldsymbol{c}_{1} \widehat{\boxdot} \boldsymbol{c}_{2} - \boldsymbol{c}_{1}^{\prime} \widehat{\boxdot} \boldsymbol{c}_{2}^{\prime}) = \boldsymbol{c}_{1d} \boldsymbol{\delta}_{-,0} + \boldsymbol{c}_{2d} \boldsymbol{\delta}_{0,-} + \kappa\left(\boldsymbol{c}_{ba}*\boldsymbol{c}_{bd}^{*-}+\boldsymbol{c}_{bd}*\boldsymbol{c}_{ba}^{*-}\right)\); and \(\frac{1}{2}(\boldsymbol{c}_{1} \widehat{\boxdot} \boldsymbol{c}_{2} + \boldsymbol{c}_{1}^{\prime} \widehat{\boxdot} \boldsymbol{c}_{2}^{\prime}) = \boldsymbol{c}_{1a} \boldsymbol{\delta}_{-,0} + \boldsymbol{c}_{2a} \boldsymbol{\delta}_{0,-} +\left(\beta+\ln{\pi}\right)\boldsymbol{\delta}_{0,0} + \kappa\left(\boldsymbol{c}_{ba}*\boldsymbol{c}_{ba}^{*-}+\boldsymbol{c}_{bd}*\boldsymbol{c}_{bd}^{*-}\right)\).

Since \(\boldsymbol{c}_{1} \sim \boldsymbol{c}_{1}^{\prime}\), \(\boldsymbol{c}_{2} \sim \boldsymbol{c}_{2}^{\prime}\), there is \(\mathrm{supp}(\boldsymbol{c}_{1d}), \mathrm{supp}(\boldsymbol{c}_{2d})\subseteq\{(k,m)||k|+|m|=1\}\), \(\mathrm{supp}(\boldsymbol{c}_{1a}), \mathrm{supp}(\boldsymbol{c}_{2a})\subseteq\{(-1,1),(1,-1),(0,0)\}\). Then \(\mathrm{supp}(\boldsymbol{c}_{ba})\subseteq\{(-1,0),(0,-1)\}\), \(\mathrm{supp}(\boldsymbol{c}_{bd})\subseteq\{(0,0)\}\), so \(\mathrm{supp}(\boldsymbol{c}_{ba}*\boldsymbol{c}_{bd}^{*-}+\boldsymbol{c}_{bd}*\boldsymbol{c}_{ba}^{*-})\subseteq\{(k,m)||k|+|m|=1\}\). Besides, \(\mathrm{supp}(\boldsymbol{c}_{1d} \boldsymbol{\delta}_{-,0} + \boldsymbol{c}_{2d} \boldsymbol{\delta}_{0,-})\subseteq\{(k,m)||k|+|m|=1\}\). Therefore, \(\mathrm{supp}(\boldsymbol{c}_{1} \widehat{\boxdot} \boldsymbol{c}_{2} - \boldsymbol{c}_{1}^{\prime} \widehat{\boxdot} \boldsymbol{c}_{2}^{\prime})\subseteq\{(k,m)||k|+|m|=1\}\).
Moreover, \(\mathrm{supp}(\boldsymbol{c}_{ba}*\boldsymbol{c}_{ba}^{*-}+\boldsymbol{c}_{bd}*\boldsymbol{c}_{bd}^{*-})\subseteq\{(-1,1),(1,-1),(0,0)\}\), and \(\mathrm{supp}(\boldsymbol{c}_{1a} \boldsymbol{\delta}_{-,0} + \boldsymbol{c}_{2a} \boldsymbol{\delta}_{0,-})\subseteq\{(0,0)\}\). Therefore, \(\mathrm{supp}(\boldsymbol{c}_{1} \widehat{\boxdot} \boldsymbol{c}_{2} + \boldsymbol{c}_{1}^{\prime} \widehat{\boxdot} \boldsymbol{c}_{2}^{\prime})\subseteq\{(-1,1),(1,-1),(0,0)\}\). Namely, \(\boldsymbol{c}_1\widehat{\boxdot}\boldsymbol{c}_2 \sim \boldsymbol{c}_1^{\prime}\widehat{\boxdot}\boldsymbol{c}_2^{\prime}\).
\(\hfill \blacksquare\)

With Lemma 2, 3, it is sufficient to prove Proposition 3. We prove Proposition 3 by mathematical induction to \(N\):

\textit{Base case:} For \(N=1\), (\ref{eq:chSp_ele}) implies \(\mathrm{supp}\left(\boldsymbol{c}_{1}^{(0)}\left( y \middle| x \right)\right) \subseteq \overline{\mathcal{K}}\), \(\boldsymbol{c}_{1}^{(0)}\left( y \middle| 0 \right) \sim \boldsymbol{c}_{1}^{(0)}\left( y \middle| 1 \right)\).

\textit{Induction step:} For each \(N \geq 2\), assume that \(\mathrm{supp}\left( \widehat{\boldsymbol{c}}_{N}^{(N-1)} \left( y_{0}^{N - 1},u_{0}^{N-2} \middle| u_{N-1} \right) \right) \subseteq \overline{\mathcal{K}}\), \( \widehat{\boldsymbol{c}}_{N}^{(N-1)} \left( y_{0}^{N - 1},u_{0}^{N-2} \middle| 0 \right) \sim \widehat{\boldsymbol{c}}_{N}^{(N-1)} \left( y_{0}^{N - 1},u_{0}^{N-2} \middle| 1 \right) \), and \(\forall i<N-1\), \(\mathrm{supp}\left( \widehat{\boldsymbol{c}}_{N}^{(i)} \left( y_{0}^{N - 1},u_{0}^{i - 1} \middle| u_{i} \right) \right) \subseteq \widetilde{\mathcal{K}}\), then:

According to (\ref{eq:aprxSRe_cmp})(\ref{eq:aprxSRo_cmp}) and Lemma 2, 3, taking \(i=N-1\), there is \(\mathrm{supp}\left(\widehat{\boldsymbol{c}}_{2N}^{(2N - 1)}\left( y_{0}^{2N - 1},u_{0}^{2N-2} \middle| u_{2N - 1} \right)\right) \subseteq \overline{\mathcal{K}}\), \( \widehat{\boldsymbol{c}}_{2N}^{(2N - 1)}\left( y_{0}^{2N - 1},u_{0}^{2N-2} \middle| 0 \right) \sim \widehat{\boldsymbol{c}}_{2N}^{(2N - 1)}\left( y_{0}^{2N - 1},u_{0}^{2N-2} \middle| 1 \right)\), \(\mathrm{supp}\left(\widehat{\boldsymbol{c}}_{2N}^{(2N - 2)}\left( y_{0}^{2N - 1},u_{0}^{2N-3} \middle| u_{2N - 2} \right)\right) \subseteq \widetilde{\mathcal{K}}\); and taking \(i<N-1\), there is \(\mathrm{supp}\left(\widehat{\boldsymbol{c}}_{2N}^{(2i + 1)}\left( y_{0}^{2N - 1},u_{0}^{2i} \middle| u_{2i + 1} \right)\right) \subseteq \widetilde{\mathcal{K}}\), \(\mathrm{supp}\left(\widehat{\boldsymbol{c}}_{2N}^{(2i + 1)}\left( y_{0}^{2N - 1},u_{0}^{2i} \middle| u_{2i + 1} \right)\right) \subseteq \widetilde{\mathcal{K}}\). Namely, \(\mathrm{supp}\left( \widehat{\boldsymbol{c}}_{2N}^{(2N-1)} \left( y_{0}^{2N - 1},u_{0}^{2N-2} \middle| u_{2N-1} \right) \right) \subseteq \overline{\mathcal{K}}\), \( \widehat{\boldsymbol{c}}_{2N}^{(2N - 1)}\left( y_{0}^{2N - 1},u_{0}^{2N-2} \middle| 0 \right) \sim \widehat{\boldsymbol{c}}_{2N}^{(2N - 1)}\left( y_{0}^{2N - 1},u_{0}^{2N-2} \middle| 1 \right)\), and \(\forall i<2N-1\), \(\mathrm{supp}\left( \widehat{\boldsymbol{c}}_{2N}^{(i)} \left( y_{0}^{2N - 1},u_{0}^{i - 1} \middle| u_{i} \right) \right) \subseteq \widetilde{\mathcal{K}}\).

\textit{Conclusion:} Since both the base case and the induction step have been proved as true, we know at each \(N\in\{2^n|n\in\mathbb{N}\}\), \(\mathrm{supp}\left( \widehat{\boldsymbol{c}}_{N}^{(N-1)} \left( y_{0}^{N - 1},u_{0}^{N-2} \middle| u_{N-1} \right) \right) \subseteq \overline{\mathcal{K}}\), \( \widehat{\boldsymbol{c}}_{N}^{(N-1)} \left( y_{0}^{N - 1},u_{0}^{N-2} \middle| 0 \right) \sim \widehat{\boldsymbol{c}}_{N}^{(N-1)} \left( y_{0}^{N - 1},u_{0}^{N-2} \middle| 1 \right) \), and \(\forall i<N-1\), \(\mathrm{supp}\left( \widehat{\boldsymbol{c}}_{N}^{(i)} \left( y_{0}^{N - 1},u_{0}^{i - 1} \middle| u_{i} \right) \right) \subseteq \widetilde{\mathcal{K}}\).
\(\hfill \blacksquare\)

\end{document}